\documentclass[a4paper,11pt]{article}
\usepackage{jcappub} 
\usepackage{orcidlink}
\usepackage{lineno}
\usepackage{amsmath}
\usepackage{amssymb}
\usepackage{graphicx}
\usepackage{subfigure}
\usepackage{color}
\usepackage{ulem}
\usepackage{pifont}
\usepackage{makecell}
\usepackage{float}
\allowdisplaybreaks

\usepackage{bm}
\usepackage{hyperref}

\title{Superhorizon curvature perturbations in hybrid inflation revisited}

\author{Shi Pi${}^{1,2}$
\,\orcidlink{0000-0002-8776-8906}}
\emailAdd{shi.pi@itp.ac.cn} 
\author{Anxianyi Xiong${}^{1,3}$\,\orcidlink{0009-0004-0900-2907}}
\emailAdd{xionganxianyi@itp.ac.cn}

\affiliation{
    $^{1}$ Institute of Theoretical Physics, Chinese Academy of Sciences, Beijing 100190, China}
\affiliation{
    $^{2}$ Kavli Institute for the Physics and Mathematics of the Universe (WPI), The University of Tokyo, Kashiwa, Chiba 277-8583, Japan}
\affiliation{
    $^{3}$ School of Physical Sciences, University of Chinese Academy of Sciences, Beijing 100049, China}

\abstract{
We revisit cosmological perturbations in multi-field inflation using the $\delta N$ formalism. By extending the analysis to directions transverse to the inflationary trajectory, we explicitly account for the geometry of the final hypersurface. Applying this framework to hybrid inflation, we identify an enhancement mechanism of the curvature perturbation driven by the growing isocurvature perturbation due to the tachyonic instability of the waterfall field. This amplification occurs during the trajectory's turn in field space, a process qualitatively distinct from the non-attractor solution in single-field inflation models such as ultra-slow-roll. The resulting power spectrum features a broad peak with a characteristic $k^3$ infrared growth and an ultraviolet spectral tilt that uniquely determines the nonlinear parameter $f_\mathrm{NL}$ of a logarithmic non-Gaussianity, all of which are primarily governed by the waterfall dynamics. 
We found that in hybrid inflation, the sign of $f_{\rm NL}$ is fixed by tachyonic waterfall geometry and is always positive, leading to a generic enhancement of primordial black hole formation.
The enhanced curvature perturbation can simultaneously account for primordial black hole dark matter and a stochastic gravitational wave background detectable by LISA, Taiji, and TianQin.}

\begin{document}
\maketitle
%\tableofcontents
\section{Introduction}
Inflation \cite{Brout:1977ix,Starobinsky:1980te,Guth:1980zm,Sato:1981qmu,Linde:1981mu,Albrecht:1982wi} constitutes a cornerstone of the modern cosmological paradigm. It not only offers compelling solutions to the horizon, flatness, and monopole problems but also provides a natural mechanism for generating the primordial density perturbations through the quantum fluctuations of the scalar field (the inflaton) that drives the inflationary expansion \cite{Mukhanov:1981xt}.

A variety of models have been proposed to realize inflation. The simplest scenario is single canonical field slow-roll inflation, wherein the inflaton rolls gradually down its potential, leading to a sufficient number of e-folds (typically $N \sim 60$) to resolve the standard cosmological puzzles. Classic examples include chaotic inflation \cite{Linde:1983gd}, natural inflation \cite{Freese:1990rb}, and $R^2$ inflation \cite{Starobinsky:1980te}. 

On scales larger than $1\ \mathrm{Mpc}$, the curvature perturbation power spectrum is constrained by cosmic microwave background radiation (CMB) and large-scale structure (LSS) observations, which is nearly scale invariant with an amplitude of $2.1 \times 10^{-9}$ at 200 Mpc \cite{Planck:2018jri}. 
%these models typically predict an almost scale-invariant primordial curvature power spectrum, with its amplitude tightly constrained by cosmic microwave background radiation (CMB) and large-scale structure (LSS) observations to be approximately $2.1 \times 10^{-9}$ \cite{Planck:2018jri}. 
However, the amplitude of the power spectrum on small scales %(e.g. modes exiting the horizon roughly $30$ e-folds before the end of inflation) 
is much less constrained, and an enhanced power spectrum  could facilitate the production of primordial black holes (PBHs) and generate a significant stochastic gravitational wave background (SGWB) sourced by the curvature perturbation at quadratic order, both of which can be used to probe the physics of the very early universe.
There are several mechanisms to amplify the power spectrum on small scales. For instance, in single-field inflation, the amplification can be achieved if the potential has a flat plateau or near-inflection point \cite{Starobinsky:1992ts,Yokoyama:1998pt,Garcia-Bellido:2016dkw,Cheng:2016qzb,Garcia-Bellido:2017mdw,Cheng:2018yyr,Dalianis:2018frf,Tada:2019amh,Xu:2019bdp,Mishra:2019pzq,Bhaumik:2019tvl,Liu:2020oqe,Fu:2020lob,Vennin:2020kng,Ragavendra:2020sop,Gao:2021dfi,Pi:2022zxs,Cole:2022xqc,Kubota:2023ked,Fujita:2025imc,Tomberg:2025fku,Talebian:2025jeg,Escriva:2025ftp}, a bump \cite{Atal:2019cdz,Atal:2019erb,Mishra:2019pzq,Karam:2022nym,Escriva:2023uko,Briaud:2023eae,Domenech:2023dxx,Gu:2023mmd,Wang:2024wxq} (\textit{i.e.} constant-roll \cite{Kinney:2005vj,Martin:2012pe,Motohashi:2014ppa,Motohashi:2017aob,Motohashi:2017vdc,Atal:2018neu,Motohashi:2019tyj,Tomberg:2023kli,Wang:2024xdl,Inui:2024sce,Inui:2024fgk,Shimada:2024eec}), or a step \cite{Cai:2021zsp,Kawaguchi:2023mgk,Wang:2024nmd}, 
%inflaton potential—such as a step \cite{Cai:2021zsp}, a near-inflection point \cite{Pi:2022zxs,Garcia-Bellido:2017mdw}, or a steep plateau \cite{Leach:2000yw}. 
which usually contains a transient stage when the non-attractor solution dominates.

%In multi-field theory, hese multi-field constructions generally impose fewer restrictions on the inflaton’s dynamical history while remaining consistent with observation constraints on large scales.  

%Beyond linear perturbation theory, 
It is convenient to use the $\delta N$ formalism \cite{Starobinsky:1985ibc,Salopek:1990jq,Sasaki:1995aw,Wands:2000dp,Lyth:2004gb} to describe the nonlinear evolution of the cosmological perturbations on superhorizon scales. 
%It was shown that such a nonlinearity is crucial to determine the statistical properties of the curvature perturbation \cite{Pi:2022ysn,Pi:2024lsu}. 
%serves as a powerful tool for analyzing the curvature perturbations nonlinearly in such models. 
In this approach, distant Hubble patches are linked by the momentum constraints which decays rapidly on superhorizon scales, which can be seen as independent ``separate universes''. They evolve independently and obey local Friedmann equations. In any Hubble patch coarse-grained on a scale larger than $H^{-1}$, the curvature perturbation $\mathcal{R}$ at the end of inflation is identified to the difference of its total local expansion (from an initial spatially-flat hypersurface to the end of inflation) and the averaged total expansion,
%(usually a comoving slice or a uniform-density slice) equals to the final curvature perturbation. radius shrinks during inflation, the universe effectively separates into causally disconnected Hubble patches—each evolving as a "separate universe".  
%The curvature perturbation $\mathcal{R}$ is identified as the difference in the $e$-fold number between the fiducial homogeneous background and a perturbed trajectory in phase-space, from an initial spatially-flat hypersurface to a final comoving or uniform-density hypersurface, 
which is usually a function of the field perturbations and their velocities, \textit{i.e.} $\mathcal{R} = N-\langle N\rangle=\delta N(\delta\varphi,\delta\pi)$. Such a formalism is fully nonlinear, which can correctly capture the non-Gaussianity of $\mathcal{R}$ \cite{Pi:2022ysn,Pi:2024lsu}. 
For instance, in constant-roll inflation, it is shown that $\mathcal{R}$ has a nonlinear form of $\mathcal{R}=-\gamma^{-1}\ln(1-\gamma\mathcal{R}_G)$ \cite{Biagetti:2018pjj,Atal:2019cdz,Atal:2019erb,Biagetti:2021eep,Passaglia:2018ixg,Pi:2022ysn,Pi:2024lsu}, which gives the so-called exponential-tail probability density function (PDF) of $\mathbb{P}(\mathcal{R})\sim\exp(-\gamma\mathcal{R})$ \cite{Biagetti:2021eep,Pi:2022ysn,Pi:2024lsu} .
%The local expansion in these patches follows a probability distribution function (PDF), and the non-Gaussian nature of $\mathcal{R}$ significantly influences the tail of this PDF.
%This relation nonlinearly connects $\mathcal{R}$ to the field and velocity perturbations, making the formalism particularly suitable for studying non-Gaussianity. 
This is especially important for accurately calculating the PBH mass function, as PBH formation is sensitive to the number of high-density peaks, which is determined by the tail part of the PDF \cite{Atal:2019cdz,Atal:2019erb,Kitajima:2021fpq,Gow:2022jfb,Escriva:2023uko,Pi:2024ert,Young:2024jsu,Escriva:2025rja,Escriva:2025ftp,Saito:2025sny}. 

Single-field inflation with a ultra-slow-roll or constant-roll stage suffers from fune-tune problem \cite{Cole:2023wyx}, which can be alleviated in multifield models, including hybrid inflation with a waterfall stage \cite{Linde:1993cn,Murata:2025onc,Dalianis:2026tty,Clesse:2015wea,Lin:2024gar,Afzal:2024xci,Tada:2023pue}, modified gravity with a scalaron \cite{Kannike:2017bxn,Pi:2017gih,Gao:2018pvq,Cheong:2019vzl,Cheong:2020rao,Fu:2019ttf,Dalianis:2019vit,Lin:2020goi,Fu:2019vqc,Aldabergenov:2020bpt,Aldabergenov:2020yok,Yi:2020cut,Gao:2020tsa,Dalianis:2020cla,Poisson:2023tja,Choudhury:2024one,Wang:2024vfv}, the curvaton scenario \cite{Kawasaki:2012wr,Kohri:2012yw,Ando:2017veq,Ando:2018nge,Pi:2021dft,Chen:2023lou,Ferrante:2023bgz,Gow:2023zzp,Inomata:2023drn,Chen:2024pge,Tokeshi:2024kuv,Su:2025mam,Kuroda:2025coa},  post-inflationary entropy-to-curvature conversion \cite{GonzalezQuaglia:2025qem,Gorgulho:2026shk} etc. Among them, hybrid inflation provides several key advantages: it not only enlarges the viable parameter space, thereby alleviating the fine-tuning issue around the enhanced peak, but also provides a graceful exit of inflation by the waterfall dynamics. In this work, we use $\delta N$ formalism to investigate the predictions of hybrid inflation models regarding the enhancement of small-scale power spectrum, which generates PBHs and induced gravitational waves. 

This paper is organized as follows. In \autoref{deltaN}, starting from the linear $\delta N$ formalism, we introduce additional coordinates in phase-space to describe the evolution of curvature and isocurvature perturbations on different final hypersurfaces. This extended approach, referred to as the phase-space formalism of perturbations in $\delta{N}$ gauge, effectively parameterizes trajectories in the reduced two-dimensional phase-space—under the assumption that both fields follow attractor solutions. In \autoref{hybinf}, we apply this method to study the dynamics of hybrid inflation. In \autoref{pert}, we choose the final hypersurface to be the constant-$\theta$ hypersurface at the end of inflation, rather than the comoving hypersurface or the uniform-density hypersurface (although they are equivalent at the end of inflation in this context \cite{Sasaki:1998ug}), with the value of the $\varphi$ field along each trajectory on that hypersurface serving as the second field-space coordinate. This setup allows us to investigate the properties of the curvature perturbation power spectrum, and approximate its shape with a parametric form. In \autoref{mecha}, we analyze the enhancement mechanism of curvature perturbations in this model using both linear perturbation theory in multi-field inflation and the phase-space formalism. The two approaches show consistent results. Moreover, the phase-space formalism, which incorporates the full field-space coordinates, reveals how isocurvature perturbations convert into curvature perturbations before the trajectory turns—a feature that cannot be captured by the standard $\delta N$ formalism alone. We discuss the non-Gaussianity in \autoref{sec:NG}. In \autoref{sec:PBH} and \autoref{sec:IGW}, we compute the resulting primordial black hole mass function and the scalar induced gravitational wave power spectrum.

%%%%%%%%%%%%%%%%%%%%%%%%%%%%%%%%%%%%%%%%%%%%%%%%%%%%%% %%%%%%%%%%%%%%%%%%%%%%%%%%%%%%%%%%%%%%%%%%%%%%%%%%%%%% %%%%%%%%%%%%%%%%%%%%%%%%%%%%%%%%%%%%%%%%%%%%%%%%%%%%%% 
\section{Super-Horizon evolution of multi scalar fields}\label{deltaN}

\subsection{General Discussion}

The $\delta N$ formalism is a powerful and widely used analytical tool in cosmology for studying the nonlinear evolution of the comoving curvature perturbation $\zeta$ on superhorizon scales. It is based on the separate-universe approach, which assumes that on scales much larger than the Hubble radius, different spatial regions evolve independently as locally homogeneous and isotropic Friedmann-Lemaître-Robertson-Walker (FLRW) universes. In this picture, quantum fluctuations of the inflaton field that exit the Hubble horizon during inflation are treated as classical perturbations, leading to different initial conditions in each Hubble patch. The local expansion history of each patch is characterized by the number of $e$-folds, $N$, and the curvature perturbation is given by the difference between the local and a fiducial $e$-folding numbers as $\zeta = \delta N$. 

We work with the perturbed metric
\begin{align}
\text{d}s^2=a^2[-(1+2AY)\text{d}\eta^2-2BY_j\text{d}\eta\text{d}x^j+((1+2H_LY)\delta_{ij}+2H_TY_{ij})\text{d}x^i\text{d}x^j].
\end{align}
where $Y$ is the spatial scalar harmonic with the eigenvalue $k^2,\;Y_j=-k^{-1}\nabla_jY$, and $Y_{ij}=k^{-2}\nabla_i\nabla_jY+\frac{1}{3}\delta_{ij}Y$.
In the $\delta N$ gauge ($H_L'=0$), the equations of motion for the $n$-scalar-field system is \cite{Sasaki:1998ug}
\begin{align}\label{pert0i}
    &2VA=-H^2\frac{\text{d}}{\text{d}N}\bm{\varphi}\cdot\frac{\text{d}}{\text{d}N}\delta\bm{\varphi}-V_{,p}\delta\varphi^p,\\\label{pert-eom}
    &H\frac{\text{d}}{\text{d}N}\left(H\frac{\text{d}}{\text{d}N}\delta \varphi^p\right)+3H^2\frac{\text{d}}{\text{d}N}\delta\varphi^p+\frac{k^2}{a^2}\delta\varphi^p+V^{,p}_{,q}\delta\varphi^q+2V^{,p}A-H^2\frac{\text{d}}{\text{d}N}\varphi^p\frac{\text{d}}{\text{d}N}A=0,
\end{align}
where $\delta\bm{\varphi}\equiv\{\delta\varphi^p\}$ is the vector in field space, with $p,q=1,\cdots,n$ the field indices, thus $\delta\varphi^p$ represents the scalar field perturbations. On superhorizon scales, \textit{i.e.} $k^2/a^2\ll1$, Eqs. \eqref{pert0i} and \eqref{pert-eom} have the same form as the equations of motion for $\delta_\lambda\bm\varphi$, the derivative of $\bm\varphi$ with respect to parameter $\lambda^\alpha$ ($\alpha=1,\cdots,2n$),
\begin{align}\label{bgd-oi}
    &-2V\frac{\delta_\lambda H}{H}=-H^2\frac{\text{d}}{\text{d}N}\bm{\varphi}\cdot\frac{\text{d}}{\text{d}N}\delta_\lambda\bm{\varphi}-V_{,p}\delta_\lambda\varphi^p,\\\label{bgd-eom}
    &H\frac{\text{d}}{\text{d}N}\left(H\frac{\text{d}}{\text{d}N}\delta_\lambda\varphi^p\right)+3H^2\frac{\text{d}}{\text{d}N}\delta_\lambda\varphi^p+V^{,p}_{,q}\delta_\lambda\varphi^q-2V^{,p}\frac{\delta_\lambda H}{H}+H^2\frac{\text{d}}{\text{d}N}\varphi^p\frac{\text{d}}{\text{d}N}\frac{\delta_\lambda H}{H}=0.
\end{align}
Similarly, $\delta_\lambda H$ indicates $\partial H/\partial\lambda$, and the index of $\lambda^\alpha$ is suppressed for simplicity.
 
The two sets of equations describe the same phase-space, where the perturbations refer to the deviations from a fiducial point in the phase-space spanned by the fields and their velocities, $\{\varphi^p,\pi^q\equiv\text{d}\varphi^q/\text{d}N\}$ with $p,q=1,\cdots,n$. 
For convenience, let us define an auxiliary field $\phi^i\equiv\{\varphi^p,\pi^q\}$ with $i=1,\cdots,2n$, \textit{i.e.} $\phi^i\equiv\varphi^i$ when $1\leq i\leq n$, and $\phi^i\equiv\pi^{i-n}$ when $n<i\leq 2n$.
Similarly, field perturbations and velocity perturbations are denoted by ${\chi}^\alpha = \{\delta\varphi^p, \delta\pi^q\}$, with $\alpha=1,\cdots,2n$, \textit{i.e.} $\chi^\alpha\equiv\delta\varphi^\alpha$ when $1\leq \alpha\leq n$, and $\chi^\alpha\equiv\delta\pi^{\alpha-n}$ when $n<\alpha\leq 2n$.
The identity of the phase space means that the perturbations $\chi^\alpha$ are directly related to the derivatives of the background fields and velocities $\phi^i$ with respect to the parameter $\lambda^\alpha$, which implies that the solution of \eqref{pert-eom} can be expressed as a linear combination of the solutions of \eqref{bgd-eom}.

Equation \eqref{pert-eom} corresponds to the familiar representation of the phase space parameterized by the fields and their velocities, while equation \eqref{bgd-eom} describes an alternative parameterization of the same phase space in terms of coordinates $\lambda^\alpha$, where $\lambda^1 \equiv N$ implies that the time evolution of each trajectory in the phase space can be described by the $e$-folding number $N$, while the other $\lambda$ label the different trajectories. Therefore, we can use the following coordinate transformation to go back and forth between these two phase spaces,
\begin{align}\label{def of chi}
 { \chi}^i=c^\alpha\cdot\frac{\partial { \phi}^i(\lambda)}{\partial\lambda^\alpha}\equiv c^\alpha\mathcal{X}^i_\alpha,
\end{align}
where $i,\alpha=1,2,\cdots,2n$. 
In this expression, $\mathcal{X}^i_\alpha \equiv \partial \phi^i / \partial \lambda^\alpha$ are the components of the Jacobian matrix of the map $\lambda \mapsto \phi(\lambda)$. Each column $\mathcal{X}^i_\alpha$ (for fixed $\alpha$) is a vector in phase space that satisfies the background equations \eqref{bgd-eom}, thus representing a tangent vector to the family of background solutions. The coefficients $c^\alpha$ have a dual interpretation: they are both the components of the perturbation vector $\chi^i$ in the basis spanned by the vectors $\mathcal{X}^i_\alpha$ (meaning that when we express $\chi^i$ in the new coordinate system $\lambda^\alpha$, its components are $c^\alpha$) and the coefficients in the linear combination of the independent background solutions $\mathcal{X}^i_\alpha$ that reconstruct the perturbation solution $\chi^i$. Thus, equation \eqref{def of chi} can be viewed as a coordinate transformation from the parameter space $\lambda^\alpha$ to the phase space $\phi^i$, where the perturbation $\chi^i$ is the image of the vector $c^\alpha$ under the differential of this map. Provided the Jacobian matrix $\mathcal{X}^i_\alpha$ is invertible (at least locally in the region of interest), we can uniquely transform between the two descriptions.

The field perturbations on a spatially flat hypersurface similar to the Mukhanov-Sasaki variable \cite{Mukhanov:1985rz,Sasaki:1986hm} can be defined as \cite{Sasaki:1998ug} 
\begin{align}\label{def:chiF}
    \chi_F^i(N)\equiv  { \chi}^i-\phi_N^i\mathcal{R}= c^\alpha\mathcal{X}_{\alpha}^i-\phi_N^i\mathcal{R},
\end{align}
$c^\alpha$ can be solved from \eqref{def:chiF}, which, as $c^\alpha$ is a constant, only depends on the perturbations on the initial hypersurface 
\begin{align}\label{c(chi,R)}
    c^\alpha=[(\mathcal{X}^{-1})^\alpha_i(\chi_F^i+\phi_N^i\mathcal{R})]_{N_0}.
\end{align}
Taking the inner product of \eqref{def:chiF} with $\phi_N^p$, and using the condition of comoving hypersurface $\phi_N^p\chi^p=0$, we have
\begin{align}\label{cocurp}
    \mathcal{R}_c(N_f)&=-\sum_{p=1}^n\frac{\pi_p\chi^p_F}{\pi^2}
   %  &=\left(\sum_{\beta=1}^{2n}\frac{\partial N}{\partial \phi^\beta}\chi_F^\beta\right)_{N_0}-\sum_{I=1}^n\sum_{a=2}^{2n}\left(\sum_{\beta=1}^{2n}\frac{\partial\lambda^a}{\partial \phi^\beta}\chi_F^\beta\right)_{N_0}\frac{\pi_I}{\pi^2}\frac{\partial\varphi^I}{\partial\lambda^a}(N_f)\nonumber\\
   %  &+\mathcal{R}(N_f)-\sum_{I=1}^n\sum_{a=1}^{2n}\left[\sum_{\beta=1}^{2n}(\mathcal{X}^{-1})^a_\beta\frac{\partial\phi^\beta}{\partial N}\mathcal{R}\right]_{N_0}\frac{\pi_I}{\pi^2}\frac{\partial\varphi^I}{\partial\lambda^a}(N_f)\nonumber\\
   %  &\approx\left(\sum_{\beta=1}^{2n}\frac{\partial N}{\partial \phi^\beta}\chi_F^\beta\right)_{N_0}-\sum_{I=1}^n\sum_{a=2}^{2n}\left(\sum_{\beta=1}^{2n}\frac{\partial\lambda^a}{\partial \phi^\beta}\chi_F^\beta\right)_{N_0}\frac{\pi_I}{\pi^2}\frac{\partial\varphi^I}{\partial\lambda^a}(N_f)\nonumber\\
   %  &+\mathcal{R}(N_0)\left(1-\sum_{I=1}^n\sum_{a=1}^{2n}\delta^a_1\frac{\pi_I}{\pi^2}\frac{\partial\varphi^I}{\partial\lambda^a}(N_f)\right)\nonumber\\
    =\left(\sum_{\beta=1}^{2n}\frac{\partial N}{\partial \phi^\beta}\chi_F^\beta\right)_{N_0}-\sum_{I=1}^n\sum_{a=2}^{2n}\left(\sum_{\beta=1}^{2n}\frac{\partial\lambda^a}{\partial \phi^\beta}\chi_F^\beta\right)_{N_0}\frac{\pi_I}{\pi^2}\frac{\partial\varphi^I}{\partial\lambda^a}(N_f).
\end{align}
where $\pi^2\equiv\sum_p(\pi^p)^2$, and $\mathcal{R}_c(N_f)$ is the curvature perturbation on the final comoving slice.
Here we have used the relation $\partial\phi^\beta/\partial N = \mathcal{X}^\beta_1$, and adopted the approximation which the decaying mode $a^{-3}$ is neglected, giving $\mathcal{R}(N_f) = \mathcal{R}(N_0)$. Henceforth, unless otherwise specified, $\mathcal{R}$ will refer to the comoving curvature perturbation $\mathcal{R}_c$, and the subscript will be omitted for brevity. 
Accordingly, $\chi_F$ is rewritten as $\delta\phi_F$ to clarify its physical meaning as the phase-space perturbation. Note that in the derivation above, we have only used the conditions of the $\delta N$ gauge, which leaves one redundant degree of freedom \cite{Artigas:2024ajh}. This implies that at the final state $N_f$, we may appropriately choose the configuration of the final hypersurface to simplify our calculations.

In multi-field models, the curvature perturbation, also dubbed as adiabatic perturbation, is defined as the perturbation along the direction of the trajectory in field space. The non-adiabatic isocurvature perturbations are the perturbations perpendicular to the trajectory. For a system of $n$ scalar fields, this definition gives one curvature perturbation and $n-1$ isocurvature perturbations. The latter can be written as \cite{Kaiser:2012ak}
\begin{align}
    \mathcal{S}^{(m)}(N)=\sum_{I,J=m}^{m+1}\frac{\epsilon_{IJ}\pi^I\chi_F^{J}}{\pi^2}=\sum_{I,J=m}^{m+1}\sum_{a=2}^{2n}\left(\sum_{\beta=1}^{2n}\frac{\partial\lambda^a}{\partial \phi^\beta}\chi_F^\beta\right)_{N_0}\frac{\epsilon_{IJ}\pi^I}{\pi^2}\frac{\partial\varphi^{J}}{\partial\lambda^a}(N),
\end{align}
where $m = 1, \cdots, n-1$ labels the isocurvature perturbations along $(n-1)$ orthogonal directions in the sub field space perpendicular to the trajectories. It can be observed that, due to the presence of the totally antisymmetric tensor $\epsilon_{IJ}$, the isocurvature perturbation does not contain $({\partial N}/{\partial \phi^\beta}\cdot\delta\phi_F^\beta)_{N_0}$ term.

%%%%%%%%%%%%%%%%%%%%%%%%%%%%%%%%%%%%%%%%%%%%%%%%%%%%%% %%%%%%%%%%%%%%%%%%%%%%%%%%%%%%%%%%%%%%%%%%%%%%%%%%%%%% %%%%%%%%%%%%%%%%%%%%%%%%%%%%%%%%%%%%%%%%%%%%%%%%%%%%%% 
\subsection{Phase-Space Analysis}
In this section, we lay the analytical foundation of the generalized phase-space $\delta N$ formalism within linear perturbation theory on super-horizon scales. By parameterizing the field trajectories with field values and velocities at a given initial flat hypersurface, denoted as $\lambda^\alpha$, we systematically identify the adiabatic and isocurvature degrees of freedom and track their evolution up to the final hypersurface. This linear treatment clarifies how the geometry of the final hypersurface explicitly enters the matching conditions for the comoving curvature perturbation $\mathcal{R}_c$. Furthermore, by assigning specific physical interpretations to the parameters $\lambda^\alpha$, we provide a transparent dictionary that maps the formal phase-space derivatives to concrete inflationary observables, paving the way for the full nonlinear extension in the subsequent section.

\subsubsection{Single-field theory}

Liouville's theorem states that the Hamiltonian evolution preserves the phase-space volume, implying that the Hamiltonian flow defines a diffeomorphism of the two-dimensional phase space of single-field inflation. This allows us to introduce a well-defined coordinate transformation
\begin{equation}
(\varphi,\pi) \;\longrightarrow\; (N,\pi_f),
\end{equation}
where $N$ is the number of $e$-folds counted along the classical trajectory, and $\pi_f$ denotes the field momentum evaluated on a fixed final comoving hypersurface, taken to be the end of inflation.
The coordinate $N$ parametrizes motion along a given phase-space trajectory, while $\pi_f$ serves as an independent transverse label distinguishing different trajectories. Since the Hamiltonian equations of motion are deterministic and invertible, each value of $\pi_f$ on the final hypersurface uniquely specifies a classical trajectory in phase space. By evolving backward along the Hamiltonian flow, this label is unambiguously transported to earlier times.
Liouville's theorem further guarantees that the associated Jacobian of the transformation $(\varphi,\pi)\to(N,\pi_f)$ is non-vanishing throughout the evolution, ensuring that this parametrization remains globally non-degenerate. Consequently, curves of constant $\pi_f$ coincide with individual phase-space trajectories, providing a globally well-defined coordinate system for the single-field inflationary dynamics.
This construction relies crucially on the two-dimensional nature of the single-field phase space and does not straightforwardly generalize to multi-field inflation, where the phase space has higher dimensionality and a single final momentum is insufficient to uniquely label trajectories.

On super-horizon scales, each Hubble patch evolves independently, which can be treated as a separate Friedmann universe. We select a fiducial Hubble patch as a reference and fix the $\delta N$ gauge by specifying a common initial time $N=N_0$, such that all patches share the same local expansion history up to this time. The phase-space coordinates of the fiducial patch at $N_0$ are denoted by $\phi^i_0$. A perturbed Hubble patch is then characterized by an initial phase-space displacement $\delta\phi_F^i$, defining a neighboring trajectory with initial condition $\phi^i_0+\delta\phi_F^i$.

Both the fiducial and perturbed trajectories subsequently evolve according to the Hamiltonian flow until they reach a prescribed final hypersurface. In the $\delta N$ formalism, the comoving curvature perturbation is obtained by comparing the accumulated number of $e$-folds between the two trajectories from the initial slice $N_0$ to this final hypersurface. The choice of the final hypersurface is part of the gauge freedom and is typically taken to be a comoving or uniform-density slice, specified by a condition $f(\phi^i)=0$. For the perturbed trajectory, this hypersurface is generally reached at a slightly shifted time $N_f+\Delta N$, inducing a corresponding deviation in the phase-space coordinates on the final slice \cite{Cruces:2025typ}.

In the single-field case, the comoving hypersurface is uniquely defined by a constant field value,
\begin{equation}
    \varphi=\varphi_f=\text{const},
\end{equation}
while the field momentum $\pi$ evaluated on this hypersurface, denoted by $\pi_f$, serves as a label distinguishing different phase-space trajectories. In terms of the coordinate system $(N,\pi_f)$ introduced above, variations along $N$ parametrize evolution along a given trajectory, whereas variations in $\pi_f$ correspond to displacements transverse to it. This implies
\begin{equation}
    \frac{\partial\varphi}{\partial\pi_f}=0,
    \qquad
    \frac{\partial\pi}{\partial\pi_f}=1.
\end{equation}
Therefore, the curvature perturbation at time $N$ given in \eqref{cocurp} is reduced to
\begin{equation}
    \mathcal{R}_c(N)
    =
    \left(
    \frac{\partial N}{\partial \phi^\beta}
    \delta\phi_F^\beta
    \right)_{N_0},
\end{equation}
while the entropy perturbation, defined as
$\mathcal{S}(N)=\delta p/p' - \delta\rho/\rho'$
\cite{Gordon:2000hv,Leach:2000yw,Artigas:2025nbm},
takes the form
\begin{equation}
    \mathcal{S}(N)
    =
    \frac{6+\eta-2\epsilon}{3(\eta+3-2\epsilon)}
    \frac{1}{\pi(N)}
    \left(
    \frac{\partial \pi_f}{\partial \phi^\beta}
    \delta\phi_F^\beta
    \right)_{N_0},
\end{equation}
where
\begin{equation}
    \epsilon \equiv -\frac{\mathrm{d}\ln H}{\mathrm{d}N},
    \qquad
    \eta \equiv \frac{\mathrm{d}\ln\epsilon}{\mathrm{d}N},
\end{equation}
are Hubble-flow slow-roll parameters. We emphasize that even in single-field inflation, the background dynamics is generically two-dimensional in phase space, with the field momentum providing a transverse degree of freedom. While this transverse direction is rendered irrelevant by the strong attractor behavior in standard slow-roll, it remains fully dynamical during non-attractor phases such as ultra-slow-roll, and can actively source super-horizon evolution and nonlinear effects \cite{Leach:2000yw}. By contrast, when extending the phase-space formalism to multi-field scenarios, a full $2n$-dimensional treatment often introduces severe analytical complexities. In many physically motivated models, the fields rapidly reach their respective local velocity attractors after horizon exit. Therefore, as a well-justified first step to isolate the multi-field geometric effects, we can restrict our subsequent analysis to these multi-field attractor solutions, thereby effectively reducing the $2n$-dimensional phase space to an $n$-dimensional configuration space governed solely by the field values.

%%%%%%%%%%%%%%%%%%%%%%%%%%%%%%%%%%%%%%%%%%%%%%%%%%%%%%%%%%%%%
%%%%%%%%%%%%%%%%%%%%%%%%%%%%%%%%%%%%%%%%%%%%%%%%%%%%%%%%%%%%%
\subsubsection{Multi-field inflation}
In a system of $n$ scalar fields, we assume the potential is sufficiently smooth and flat so that the dynamics of the inflaton fields can be described by the slow-roll approximation
\begin{align}
    \pi^p\approx-\frac{V^{,p}}{3H^2},\quad H^2\approx\frac{1}{3M_p^2}V.
\end{align}
 Consequently, each scalar field is considered to reside on its attractor solution, where the momenta cease to be independent dynamical variables. The original $2n$-dimensional phase space then reduces to an $n$-dimensional field space. Subsequent calculations will demonstrate the validity and effectiveness of this approximation. Therefore, the following phase-space coordinate transformation can be chosen:
\begin{align}
    (\varphi^1,\varphi^a)\to(N,\lambda^a=\varphi_f^a)\quad a=2,\cdots,n
\end{align}
The final hypersurface is taken to be a comoving hypersurface, which is equivalent to a hypersurface of uniform density, comoving, or constant Hubble scale on super-horizon scales. It satisfies \cite{Sasaki:1998ug}
\begin{align}
    {{\pi}_{fI}}(\varphi^I-{\varphi}_f^I)=0\quad I=1,\cdots,n\label{condcov}
\end{align}
Here, $\pi_f$ and $\varphi_f$ represent the phase-space values at the intersection point of the perturbed trajectory and the final hypersurface defined by $f(\varphi^I(\lambda^a)) = 0$. Note that the derivatives $\partial\varphi^I/\partial\lambda^J$ are computed on the final hypersurface by definition, from which the matrix $\partial\varphi^I/\partial\lambda^J$ can be constructed as follows:
\begin{align}
    \frac{\partial\varphi^I}{\partial\lambda^J}=\begin{cases}
        \pi^I,\quad\text{for } J=1;\\ 
        \\
        \displaystyle-\frac{\pi^J}{\pi^1},\quad\text{for } I=1,\;J=2,\cdots,n;\\
        \\
        \delta^I_J,\quad\text{for } I,\,J=2,\cdots,n.
    \end{cases}   
\end{align}
 Furthermore, from the above expression we can derive
\begin{align}
    {{\pi}_{fI}}\frac{\partial\varphi^I}{\partial\lambda^a}(N_f)=0
\end{align}
The above equation can also be obtained by directly taking the partial derivative of the comoving hypersurface equation \eqref{condcov} with respect to the coordinates $\lambda^a$. From \eqref{cocurp}, it can be seen that the comoving hypersurface condition ensures that the comoving curvature perturbation does not depend on the choice of other coordinates, i.e., the terms proportional to other coordinates $\lambda^a$ vanish because of the geometric properties of the comoving hypersurface. This result is consistent with directly transforming the curvature perturbation $\mathcal{R}$ from the $\delta N$ gauge to the gauge-invariant curvature perturbation $\mathcal{R}_c$ in the comoving gauge, yielding 
\begin{align}
    \mathcal{R}_c=\Delta N
\end{align}
This is a consequence of the geometric properties of the comoving hypersurface.

\subsection{Gauge Redundancy}
In multi-field inflation, the choice of the final hypersurface is in general nontrivial, since different field directions may remain dynamically relevant. Nevertheless, in many practical analytical treatments, it is convenient to specify the final slice by fixing the value of a particular field, in close analogy with the single-field case. This choice is justified when the background dynamics approaches an attractor regime toward the end of inflation. In such situations, the field-space trajectory becomes effectively aligned with a single adiabatic direction, which may be identified with one specific field, say $\varphi^1$. As a result, fluctuations orthogonal to this direction are suppressed, and the hypersurface of constant $\varphi^1$ becomes equivalent, up to slow-roll suppressed corrections, to the comoving hypersurface. Under these conditions, the comoving curvature perturbation is fully captured by the difference in the accumulated number of $e$-folds, $\Delta N$, between neighboring trajectories, and its power spectrum can be consistently computed within the $\delta N$ formalism using this choice of final slice.

If the field-space trajectory does not approach an attractor aligned with a single field direction toward the end of inflation, or if one is interested in the time evolution of the curvature perturbation during inflation rather than only its final value, specifying the final hypersurface by fixing the value of a particular field is no longer appropriate. In such cases, the curvature perturbation cannot be fully characterized by $\Delta N$ alone, since additional phase-space directions remain dynamically relevant and contribute to the mapping between initial conditions and $\mathcal{R}_c$. From the phase-space perspective, this reflects the fact that after transforming to a coordinate system adapted to the background flow, variations orthogonal to the $N$ direction generally do not vanish on a generic equal-field hypersurface. Their contributions must therefore be taken into account when computing the curvature perturbation. In contrast, if the final hypersurface is chosen to be the comoving hypersurface at each moment by \eqref{condcov}, these additional contributions vanish identically by construction. In this case, the curvature perturbation is entirely captured by the local difference in the accumulated number of $e$-folds, and considering only $\Delta N$ remains a consistent and gauge-invariant description.

As an illustrative example, let us consider a two-field model. Rather than introducing new coordinates in field space, it is convenient to parametrize the family of classical background trajectories by a pair of variables $(N,M)$, where $N$ denotes the number of $e$-folds along a given trajectory, and $M$ labels different trajectories in the space of solutions. The field-space coordinates are then expressed as functions $\varphi^I=\varphi^I(N,M)$. In this parametrization, derivatives with respect to $N$ and $M$ correspond respectively to directions tangent and transverse to the background flow in field space. The curvature and isocurvature perturbations can thus be written as
\begin{align}
    \mathcal{R}_c(N)&=\Delta N-\Delta M\frac{1}{\pi^2}\frac{\partial \varphi^I}{\partial N}\frac{\partial\varphi_I}{\partial M}\equiv  \mathcal{R}_{f=0}(N)-\Delta M\frac{1}{\pi^2}\frac{\partial \varphi^I}{\partial N}\frac{\partial\varphi_I}{\partial M}\label{RDNDM}\\
    \mathcal{S}(N)&=\epsilon_{IJ}\frac{1}{\pi^2}\frac{\partial\varphi^I}{\partial N}\frac{\partial\varphi^J}{\partial M}\Delta M\label{SDNDM}
\end{align}
where $\pi^2 \equiv \partial_N \varphi^I \partial_N \varphi_I$, and  $\mathcal{R}_{f=0}(N)$ denotes the curvature perturbation evaluated on a generic final hypersurface specified by $f(\varphi^I)=0$, obtained from the difference in the accumulated number of $e$-folds between this hypersurface and the initial slice. The comoving curvature perturbation is related to $\mathcal{R}_{f=0}$ by
\begin{align}
    \mathcal{R}_c(N)&=\mathcal{R}_{f=0}(N)-\mathcal{S}(N)\cdot\cot[\theta(N)]\label{RST}\\
    \mathcal{S}(N)&=\frac{1}{\pi^2(N)}\det[J](N)\cdot\Delta M
\end{align}
where $\theta(N)$ is the angle between the tangent vector of the background trajectory and the final hypersurface, and $\det[J]$ denotes the Jacobian determinant associated with the parametrization $\varphi^I=\varphi^I(N,M)$. These expressions make explicit that, in multi-field inflation, the difference between $\mathcal{R}_c$ and $\mathcal{R}_{f=0}$ arises entirely from the isocurvature perturbation. The amplitude of $\mathcal{S}$ is controlled by the transverse deformation of the field-space trajectory, as quantified by the Jacobian determinant. This provides a geometric characterization of isocurvature effects in terms of the background evolution in field space.

In the nonlinear formulation of the $\delta N$ formalism \cite{Sasaki:1998ug}, the large-scale perturbations are constructed from a complete set of background solutions parametrized by constants $\{\lambda^\alpha\}$. In this framework, the nonlinear field perturbation $\chi_F^p$ can be expressed as a linear combination of the variations of these parameters
\begin{align}
    \chi_F^p = c^\alpha\,\frac{\partial\varphi^p(N,\lambda)}{\partial\lambda^\alpha},
\end{align}
as is shown in \eqref{def of chi}. The coefficients $c^\alpha$ represent the components of the perturbation in the basis defined by the nonlinear coordinate transformation from parameter space to field configuration space. Equivalently, in a more compact notation one can write
\begin{align}
    (\mathcal{X}^{-1})^\alpha{}_p\,\chi_F^p \;=\; \Delta\lambda^\alpha,
\end{align}
where $(\mathcal{X}^{-1})^\alpha{}_p$ is the inverse of the transformation matrix between the field perturbations $\chi_F^p$ and the parameter variations $\Delta\lambda^\alpha$. This relation makes explicit that, at the nonlinear level, the curvature and isocurvature perturbations are tied to changes in the underlying background solution parameters. Consistent specification of a final hypersurface (comoving or uniform density) must also be imposed in terms of the full nonlinear solution family. 

 Here the energy density $\rho$ and the velocity potential are given by the energy-momentum tensor $T^\mu_\nu$.
 \begin{align}
     \partial_i q = T^0_i,\quad \rho=-T^0_0\nonumber.
 \end{align}
 thus the hypersurface condition must be written as
 \begin{align}
    q(N,\lambda^a)=q_{\rm ref}\qquad\text{or}\qquad \rho(N,\lambda^a)=\rho_{\rm ref},
 \end{align}
This ensures that the $\delta N$ formula remains valid beyond the linear regime and properly accounts for all nonlinear effects in multi-field dynamics.

%%%%%%%%%%%%%%%%%%%%%%%%%%%%%%%%%%%%%%%%%%%%%%%%%%%%%%
%%%%%%%%%%%%%%%%%%%%%%%%%%%%%%%%%%%%%%%%%%%%%%%%%%%%%%
%%%%%%%%%%%%%%%%%%%%%%%%%%%%%%%%%%%%%%%%%%%%%%%%%%%%%%
\section{A Case Study: Hybrid Inflation}\label{hybinf}

Hybrid inflation is a typical two-field inflationary scenario originally proposed to address graceful exit and spontaneous symmetry breaking in inflationary dynamics \cite{Linde:1991km,Linde:1993cn,Liddle:1993fq,Mollerach:1993sy,Copeland:1994vg}. In its simplest realization, the inflationary potential involves an inflaton field $\varphi$ coupled to a secondary “waterfall” field $\theta$. During the early stage of inflation, the inflaton slowly rolls down a relatively flat direction of the potential, sustaining an extended period of quasi–de Sitter expansion. The coupling to the waterfall field induces a large effective mass for $\theta$ that stabilizes it near a local minimum, thereby suppressing its dynamics. When the inflaton reaches a critical field value $\varphi_c$, the effective mass-squared of the waterfall field becomes negative, signaling a tachyonic instability and triggering a turn in the moduli space. After this so-called waterfall transition, inflation continues in the $\theta$ direction like a small-field inflation, which ends around the global minimum of $\theta$ where $\theta$ starts to oscillate \cite{Kodama:2011vs}. Hybrid inflation naturally interpolates between two slow-roll stages in different directions, while the intermediate waterfall transition stage can be either slow-roll or fast-roll, depending on the parameters. The evolution of perturbations, however, does not follow such a clear stage separation as in the background dynamics, and perturbations in both fields play crucial roles throughout both stages.

If the waterfall field possesses a $\mathbb{Z}_2$ symmetry, its background evolution near the origin can be significantly influenced by perturbations \cite{Clesse:2010iz,Clesse:2013jra,Clesse:2015wea}, potentially leading to domain wall formation \cite{Assyyaee:2015ale}. To avoid such complications, we choose to slightly break this symmetry by adding a tiny bias term, which ensures the waterfall field to roll down along a fixed trajectory from the beginning \cite{Braglia:2022phb,Wang:2024vfv}. In practice, the bias should be sufficiently large such that the classical displacement of the waterfall field dominates over its quantum fluctuations,
\begin{align}
    |\Delta \theta_{\rm cl}| > \frac{H}{2\pi},
\end{align}
thereby preventing different Hubble patches from evolving toward opposite vacua.

The potential of our model is given by
\begin{align}
    V(\theta,\varphi)&=V_0\left(1-\frac{M^2}{4V_0}\theta^2\right)^2+\frac{1}{2}m^2\varphi^2+\frac{1}{2}M^2\left(\frac{\varphi}{\varphi_c}\right)^2(\theta-\alpha\varphi_c)^2.\label{potential}
\end{align}
Where $\varphi_c$ is the value of $\varphi$ when the effective mass-square of $\theta$ becomes negative. For convenience, we redefine the following dimensionless fields
\begin{align}
    \xi=\frac{\varphi}{\varphi_c},\quad \chi=\frac{\theta}{\varphi_c},
\end{align}
for which the equations of motion are
\begin{align}\label{eomxi}
      \frac{1}{3-\epsilon}\frac{\text{d}^2\xi}{{\text{d}N^2}}+ \frac{\text{d}\xi}{\text{d}N}&=\eta_\theta\left[\beta^2+(\chi-\alpha)^2\right]\xi,\\\label{eomchi}
\frac{1}{3-\epsilon}\frac{\text{d}^2{\chi}}{{\text{d}N^2}}+\frac{\text{d}\chi}{\text{d}N}&=\eta_\theta\left[-1+\xi^2\left(1-\frac\alpha\chi\right)\right]\chi, 
\end{align}
where $\beta\equiv m/M$, and the slow roll parameters are \cite{Wands:2002bn}
\begin{align}\label{def:epsilon}
    \epsilon&=\frac{1}{2M_p^2}(\varphi'^2+\theta'^2)=\epsilon_\varphi+\epsilon_\theta\approx\frac{M_p^2}{2}\left(\frac{V_{\varphi}}{V}\right)^2+\frac{M_p^2}{2}\left(\frac{V_{\theta}}{V}\right)^2,\\\label{def:etaphi}
        % \eta_{IJ}&=\frac{1}{V}\frac{\partial^2 V}{\partial\phi^I\partial\phi^J}\equiv\frac{V_{IJ}}{V},\qquad M_p=1\\
    \eta_\varphi&\equiv M_p^2\frac{V_{\varphi\varphi}}{V}\approx M_p^2\frac{m^2}{V_0},\\\label{def:etatheta}
    \eta_\theta&\equiv M_p^2\frac{V_{{\theta\theta}}}{V}\approx-M_p^2\frac{M^2}{V_0}.
    % ,\quad \eta_\varphi=-\beta^2\eta_\theta
\end{align}
The slow-roll conditions break down when $\epsilon_\theta = 1$ for $I=\varphi,\theta$ , which means
\begin{align}
    \epsilon_\theta = \frac{M_p^2}{2}\left[\frac{M^2\theta}{V_0(1-M^2\theta^2/(4V_0))}\right]^2=1,
\end{align}
we obtain
\begin{align}
    \chi_f=(\varphi_c/M_p)^{-1}\cdot2[\sqrt{|\eta_\theta|+\eta_\theta^2/2}-|\eta_\theta|/\sqrt{2}]/|\eta_\theta|.
\end{align}
which denotes the end point of inflation.

Finally, the bias term should be large enough to make the classical drift of
the waterfall field dominate over its quantum fluctuation.  Around the initial
trajectory $\theta_c=\alpha\varphi_c$, the classical change of $\theta$ within
one $e$-fold is estimated as
\begin{align}
    |\Delta\theta_{\rm cl}|
    \simeq
    \left|\frac{V_\theta}{3H^2}\right|
    \simeq
    \left|\frac{M^2}{3H^2}\alpha\varphi_c\right|
    =
    |\eta_\theta|\alpha\varphi_c .
\end{align}
Therefore, our parameter choice should satisfy
\begin{align}\label{eq:alphaconstraint}
    |\eta_\theta|\alpha\varphi_c
 >
    \frac{H}{2\pi},
    \qquad
    \text{or equivalently}
    \qquad
    \alpha
    >
    \frac{H}{2\pi |\eta_\theta|\varphi_c}.
\end{align}
This condition ensures that the waterfall field follows the biased classical
trajectory instead of being dominated by quantum diffusion.

%%%%%%%%%%%%%%%%%%%%%%%%%%%%%%%%%%%%%%%%%%%%%%%%%%%%%% %%%%%%%%%%%%%%%%%%%%%%%%%%%%%%%%%%%%%%%%%%%%%%%%%%%%%% %%%%%%%%%%%%%%%%%%%%%%%%%%%%%%%%%%%%%%%%%%%%%%%%%%%%%% 
\subsection{Dynamics}
The dynamical evolution has  two different stages. In the beginning, the inflaton rolls down along $\varphi$ direction from a large initial value, where the effective mass of the scalar field $\theta$ is large, guaranteeing that $\theta$ stays at its stable point $\theta = \alpha \varphi_c$. During this stage, the model effectively reduces to a single-field inflation, or more precisely the \textit{false-vacuum inflation} with a potential of $V_0+(1/2)m^2\varphi^2$ \cite{Copeland:1994vg}. 
As $\varphi$ rolls down the potential gradually (but still remains greater than the critical value $\varphi_c$), the stable point of $\theta$ undergoes slight variations in response to the evolution of $\varphi$. As will be shown later, this variation contributes to $\Delta N$. Until $\varphi$ reaches its critical value, the dynamics is similar to that of a single-field inflation with a potential of $V(\varphi) = V_0 + \frac{1}{2}m^2\varphi^2$, which helps us to solve the slow-roll equation of motion, and get the the $e$-folding number
\begin{align}
    \pi=\frac{-2\eta_\varphi\varphi}{\eta_\varphi(\frac{\varphi}{M_p})^2+2}.
\end{align}
From this, the $e$-folding number is derived as follows
\begin{align}
    N_c-N&=\frac{\varphi^2-\varphi_c^2}{4M_p^2}+\frac{1}{\eta_\varphi}\ln\left(\frac{\varphi}{\varphi_c}\right)=\frac{1}{4}\left(\frac{\varphi_c}{M_p}\right)^2(\xi^2-\xi_c^2)+\frac{1}{\eta_\varphi}\ln\left(\frac{\xi}{\xi_c}\right).\label{S01N}
\end{align}
Here, $N_c$ denotes the $e$-folding number at the end of Stage-1, which also marks the moment when the $\varphi$ field reaches the critical point. 
\begin{figure}[htbp]
\centering
    \includegraphics[width=1\textwidth]{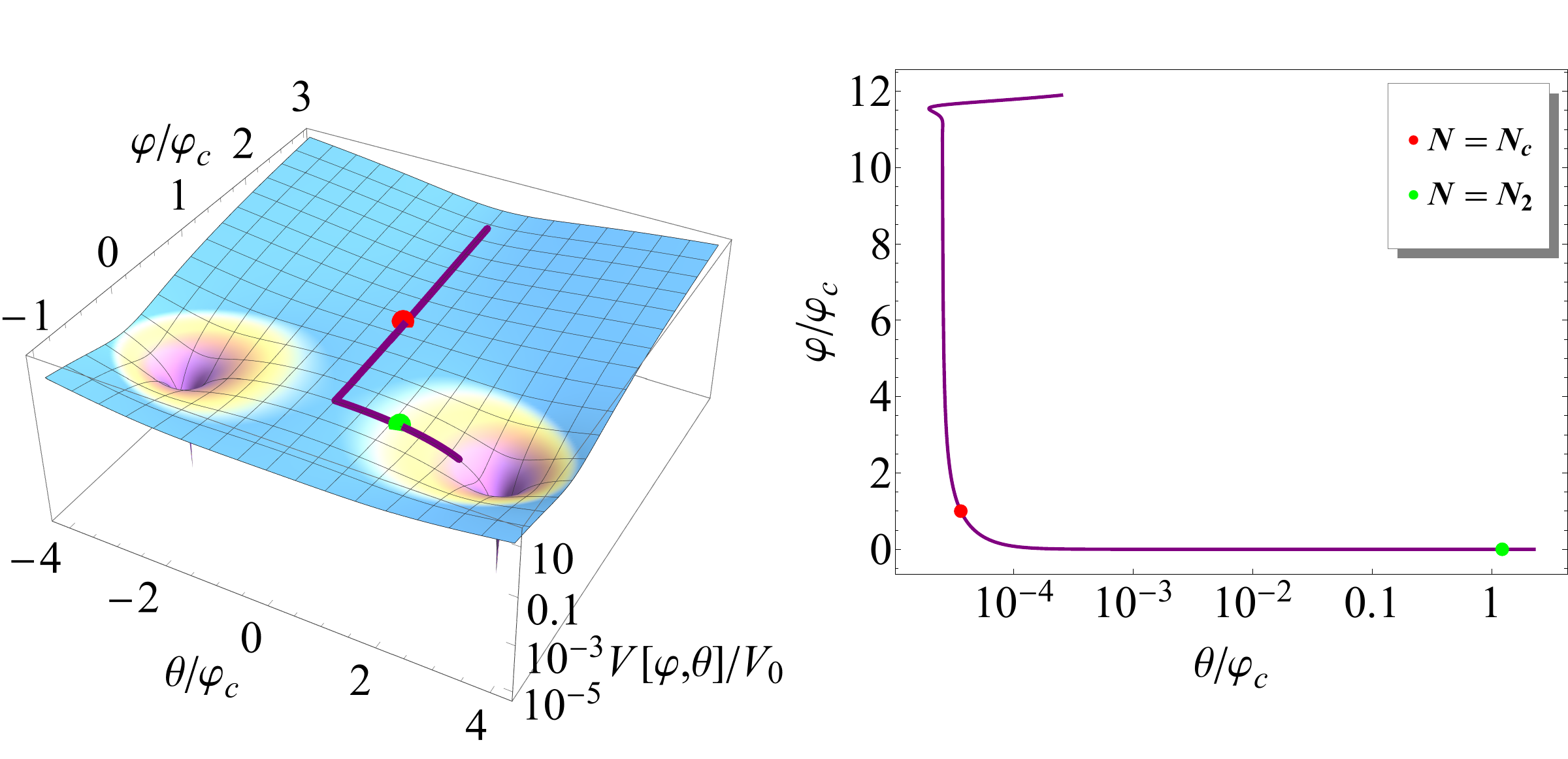}
  \caption{The left and right panels show the potential shape and the field-space trajectory. Along the trajectory, the red and green points (left to right) indicate the critical point (zero effective mass for the $\theta$ field) and the end of inflation, respectively.}
    \label{ftr}
\end{figure}

%\textbf{In Stage-2}
In the second stage, the value of $\varphi$ falls below the critical value $\varphi_c$. At this point, the effective mass-square of the scalar field $\theta$ becomes negative, and it begins to roll far away from its previously stable point.  Since the original equations of motion \eqref{eomxi} and \eqref{eomchi} contain coupling terms that make them difficult to solve directly, we adopt a perturbative approximation. We assume $(\chi-\alpha)^2\ll\beta^2$. Shortly after the critical point, once the waterfall field has grown sufficiently such that $\chi\gg\alpha$, the bias-induced source term proportional to $\alpha\xi^2$ can be neglected. We then treat the coupling term $\xi^2\chi$ perturbatively.. This leads to
\begin{align}
    \xi''+3\xi'+3\eta_\varphi\xi&=0,\\ \chi''+3\chi'+3\eta_\theta\chi&=3\eta_\theta\xi^2\chi.
    %\quad \chi=\chi^{(0)}+\chi^{(1)}+\cdots
\end{align}
At leading order, the dynamics of $\xi$ and $\chi^{(0)}$ are equivalent to that of a constant-roll stage \cite{Motohashi:2014ppa,Inui:2024sce}, with the solution:
\begin{align}
\xi&=C_1\text{e}^{-\lambda_\varphi N}+C_1'\text{e}^{-\lambda'_\varphi N},\\
\chi^{(0)}&=C_2\text{e}^{-\lambda_\theta N}+C_2'\text{e}^{-\lambda'_\theta N},
\end{align}
where $C, C'$ are coefficients determined by the initial conditions. Here $\lambda_X$ and $\lambda_X'$ are defined as the characteristic roots of $X''-3X'+3\eta_XX=0$ with $\lambda_X<\lambda_X'$ \cite{Pi:2022ysn}. For simplicity, we only consider the case that each scalar field is in the attractor, hence $C_1' = C_2' = 0$. Therefore, up to $\chi^{(1)}$, we can write the equations of motion for $\xi$ and $\chi$ as
\begin{align}\label{eom:xi}
     \frac{\text{d}\xi}{\text{d}N}&=\lambda_\theta\beta_e^2\xi=-\lambda_\varphi\xi,\qquad
 \frac{\text{d}\chi}{\text{d}N}=\lambda_\theta\left(-1+s\xi^2\right)\chi.
\end{align}
Here the parameters are
\begin{align}
  \lambda_{\varphi}&=  \frac{3-\sqrt{9-12\eta_{\varphi}}}{2},
  \quad\quad\quad\lambda_\theta= \frac{3-\sqrt{9-12\eta_{\theta}}}{2},\\
  \beta_e^2&=-\frac{3-\sqrt{9-12\eta_\varphi}}{3-\sqrt{9-12\eta_\theta}},\quad\quad
  s=\frac{3-\lambda_\theta}{3-2\lambda_\theta-2\lambda_\varphi}.
\end{align}
By dividing the two equations above, we can derive the relation between the final field values $\xi_2$, $\chi_2$ at the end of Stage-2 and their initial values $\xi$, $\chi$
\begin{align}
\xi_2^2=-\frac{1}{s}W_0[Y(\xi,\chi,\chi_2)]\qquad Y(\xi,\chi,\chi_2)=-s\exp\left[-s\xi^2+\ln\xi^2+2\beta_e^2\ln\left(\frac{\chi}{\chi_2}\right)\right],\label{xi2}
\end{align}
where $W_0[Y]$  denotes the principal branch of the Lambert $W$ function. 
A simple formula of $e$-folding number
\begin{align}
    N_2-N_{\rm ini}=\frac{1}{\lambda_\varphi}\ln\left(\frac{\xi}{\xi_2}\right)\label{N2}
\end{align}
can be derived from the equation of motion for $\xi$ \eqref{eom:xi} directly.

%%%%%%%%%%%%%%%%%%%%%%%%%%%%%%%%%%%%%%%%%%%%%%%%%%%%%% %%%%%%%%%%%%%%%%%%%%%%%%%%%%%%%%%%%%%%%%%%%%%%%%%%%%%% %%%%%%%%%%%%%%%%%%%%%%%%%%%%%%%%%%%%%%%%%%%%%%%%%%%%%% 
\subsection{Perturbation}\label{pert}
We use phase-space formalism to calculate the curvature perturbation power spectrum. So we should consider the perturbation modes exit the horizon in each stage. Indeed, when calculating the curvature perturbation power spectrum, the phase-space formalism is equivalent to the $\delta N$ formalism, except that it does not require a strict choice of the comoving hypersurface as the final hypersurface. This flexibility simplifies our calculations.
\subsubsection{Mode exits in Stage-2}
The modes exit in stage-2 contribute the peak of the curvature perturbation power spectrum. By using the formula of $e$-folding number \eqref{N2} and the finial value of $\xi$ \eqref{xi2}, we can obtain
\begin{align}
     N_2-N_{\rm ini}=\frac{s}{2\lambda_\varphi}\xi^2+\frac{1}{\lambda_\theta}\ln\left(\frac{\chi}{\chi_2}\right)+\frac{1}{2\lambda_\varphi}W_0[Y(\xi,\chi,\chi_2)]+\mathrm{const}.\label{N}
\end{align}
At this point, $N_2$ can be approximately regarded as the end of inflation, leading to the estimate $\chi_2 = \chi_f$. Moreover, we have chosen the final hypersurface for computing the power spectrum to be the constant $\chi$ hypersurface. Since $\xi_2 \ll 1$, it follows that $W_0[Y] \ll 1$. We can therefore use the expansion of the Lambert $W$ function near zero, $W_0[Y] \approx Y - Y^2 + O(Y^3)$, to obtain
\begin{align}
     N\equiv N_2-N_{\rm ini}\approx\frac{s}{2\lambda_\varphi}\xi^2\left(1-\exp\left[-s\xi^2+2\beta_e^2\ln\left(\frac{\chi}{\chi_2}\right)\right]\right)+\frac{1}{\lambda_\theta}\ln\left(\frac{\chi}{\chi_2}\right).
\end{align}
The comoving curvature perturbation $\mathcal{R}_c = \Delta N$ admits the following expansion:
\begin{align}
    \mathcal{R}_c=N_I\delta\phi^I+\frac{1}{2}N_{IJ}\delta\phi^I\delta\phi^J+\cdots
\end{align}
We can then derive the nonlinear parameter
\begin{align}
    f_{NL}=\frac{5}{6}\frac{N_{IJ}N_IN_J}{(N_K^2)^2}=-\frac{5}{6}\lambda_\theta\left[1-s(1+2\beta_e^2)\xi^2\left(\frac{\chi}{\chi_2}\right)^{2\beta^2_e}+O\left(\xi^4,\left(\frac{\chi}{\chi_2}\right)^{4\beta^2_e}\right)\right],\label{fnl}
\end{align}
which is consistent with leading order expansion from a general formula derived as in Ref. \cite{Pi:2022ysn}.
The power spectrum is
\begin{align}
   & \mathcal{P}_{\mathcal{R}}(k)=\gamma^2\frac{H_k^2}{4\pi^2}(N_\varphi^2+N_\theta^2)=\gamma^2\frac{V_0}{12\pi^2\lambda_\theta^2\theta_k^2M_p^2}\left[1+2s\left(\frac{\varphi_k}{\varphi_c}\right)^2\left(\frac{\theta_k}{\theta_2}\right)^{2\beta_e^2}+\cdots\right],\label{PS}\\
    &\gamma^2=\Gamma^2(\nu)\frac{2^{2\nu}}{2\pi},
    \quad\quad
    \nu=\frac{3}{2}-\lambda_\theta,
\end{align}
where the higher order terms in \eqref{PS} are $O\left((\varphi_k/\varphi_c)^4\right)$ and $O\left((\theta_k/\theta_2)^{4\beta_e^2}\right)$. During Stage-2, as $\theta$ gradually increases, there is a peak in the power spectrum near the critical value $\theta_c$.
\begin{figure}[htbp]
    \centering
    \includegraphics[width=0.8\textwidth]{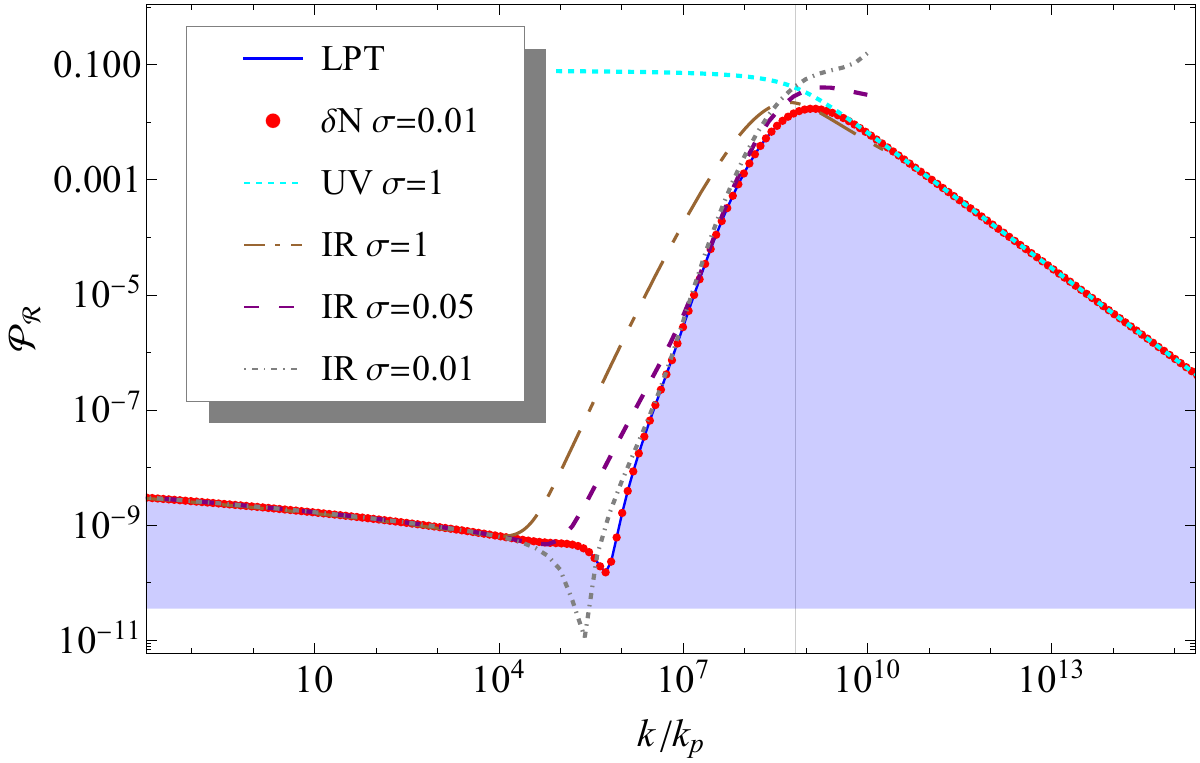} % 第二张图片
    \caption{The power spectrum of the curvature perturbation calculated at different gradient expansion parameter $\sigma$. Here $k_c$ denotes the wavenumber of the perturbation that exits the horizon at the critical point. The solid blue line represents the curvature perturbation power spectrum obtained from linear perturbation theory (LPT), the red dots denote the results from the $\delta N$ formalism, and the dashed lines show the analytical results from the $\delta N$ formalism for the IR modes with $k<k_c$ and the UV modes with $k>k_c$ under different gradient expansion parameters $\sigma$.
    }
    \label{PR}
\end{figure}

From \eqref{N} and \eqref{fnl}, we find that in hybrid inflation, a logarithmic relation $\mathcal{R}_c = ({1}/\lambda_\theta)\ln\left(1 + {\delta\chi}/{\chi}\right)$ is dominant near the peak of the power spectrum, with a nonlinear parameter  $f_{{NL}} = -{5}\lambda_\theta/{6}$, which closely resembles the result in single-field inflation. This suggests that the logarithmic behavior near the peak is primarily governed by the $\chi$ field, i.e., the waterfall field.

In our multi-field hybrid inflation model, the peak of the curvature power spectrum is reached well after the trajectory turn in field space. Consequently, the statistical properties of the curvature perturbation near this peak are entirely dictated by the waterfall field and its subsequent dynamics. Because the waterfall phase rapidly converges to a localized attractor solution, this post-turn evolution is uniquely governed by the shape of the potential, as will be demonstrated explicitly in the following subsection. This behavior is physically reminiscent of the smooth transition from ultra-slow-roll to slow-roll inflation in single-field models, where the subsequent attractor phase effectively washes out the large non-Gaussianity generated during the preceding non-attractor plateau \cite{Cai:2018dkf,Pi:2022ysn}. 

To formally implement this waterfall-dominated dynamics, we choose a final constant-field hypersurface $\chi = \chi_f$. Under the condition $|h| \ll 1$, where $h \equiv -6\sqrt{\epsilon_{\theta c}/\epsilon_{\varphi c}}$ characterizes the steepness of the critical point, the inflationary trajectory asymptotically approaches a single-field attractor before ending. This cross-section naturally coincides with the comoving hypersurface, thereby rigorously justifying the generalized $\delta N$ relation $\mathcal{R}_c = \Delta N$. To capture the full non-linear evolution, we analytically express the final phase-space coordinate $\xi_2$ on this hypersurface using the Lambert $W$ function instead of elementary field values.

%%%%%%%%%%%%%%%%%%%%%%%%%%%%%%%%%%
%%%%%%%%%%%%%%%%%%%%%%%%%%%%%%%%%%
%%%%%%%%%%%%%%%%%%%%%%%%%%%%%%%%%%

\subsubsection{Mode exits the horizon in Stage-1}
If we simply consider the perturbation of $e$-folding number in Eq. \eqref{S01N}, we obtain only a nearly scale-invariant spectrum, as the information of earlier evolution is absent. This implies that the dynamics of $\theta$ field is crucial for generating the enhancement in the power spectrum. Therefore, we start from the coupled equations of motion
\begin{align}
   \frac{\text{d}\xi}{\text{d}N}&=\frac{-2\eta_\varphi\xi}{\eta_\varphi({\varphi_c}/{M_p})^2\xi^2+2},\\
   \frac{\text{d}\chi}{\text{d}N}&=\frac{2\eta_\theta[\xi^2(\chi-\alpha)-\chi]}{\eta_\varphi({\varphi_c}/{M_p})^2\xi^2+2},
\end{align}
which gives
\begin{align}
    \chi=&\text{e}^{\frac{\xi^2}{2\beta^2}}\bigg{\{}\alpha \left(\frac{\xi^2}{2\beta^2}\right)^{\frac{-1}{2\beta^2}}\Gamma\left(1+\frac{1}{2\beta^2},\frac{\xi^2}{2\beta^2}\right)\nonumber\\
    &+\left(\frac{\xi_c}{\xi}\right)^{\frac{1}{\beta^2}}\left[\text{e}^{-\frac{\xi_c^2}{2\beta^2}}\chi_c-\alpha \left(\frac{\xi_c^2}{2\beta^2}\right)^{\frac{-1}{2\beta^2}}\Gamma\left(1+\frac{1}{2\beta^2},\frac{\xi_c^2}{2\beta^2}\right)\right]\bigg{\}},\label{chiev}
\end{align}
where $\Gamma[a,z]$ denotes the incomplete Gamma function $ \Gamma(a,z)=\int^\infty_z t^{a-1}\text{e}^{-t}\text{d}t\nonumber$. 
\begin{figure}[H]
\centering
    \includegraphics[width=0.8\textwidth]{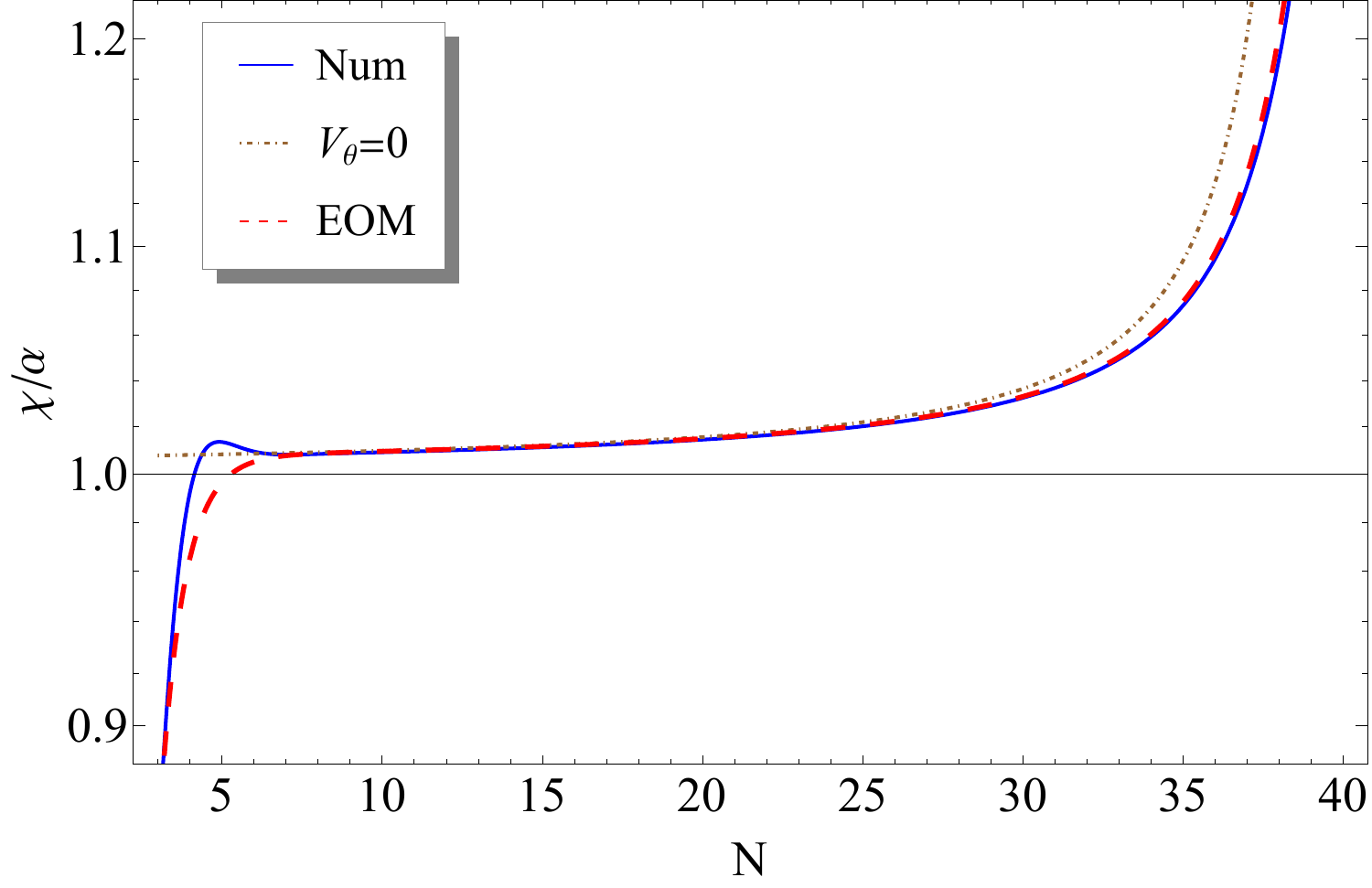}
  \caption{A slight deviation of the $\chi$ from its stable point. The brown dash-dotted line is obtained by solving for the local minimum of the $\theta$-field potential; the red dashed line represents the analytical approximate solution\eqref{chiev} derived from the equations of motion; and the blue solid line corresponds to the numerical solution.
  }
    \label{chi}
\end{figure}

Consequently, $\chi_c$ can be expressed as a function of $\chi,\xi$ and $\xi_c$, since at this stage the perturbation in $\chi_c$ is evidently larger than that in $\xi_c$ (where $\xi_c\approx1$). This yields
\begin{align}
\chi_c=&\text{e}^{\frac{\xi_c^2}{2\beta^2}}\bigg{\{}\alpha \left(\frac{\xi_c^2}{2\beta^2}\right)^{-\frac{1}{2\beta^2}}\Gamma\left(1+\frac{1}{2\beta^2},\frac{\xi_c^2}{2\beta^2}\right)\nonumber\\
&+\left(\frac{\xi}{\xi_c}\right)^{\frac{1}{\beta^2}}\left[\text{e}^{-\frac{\xi^2}{2\beta^2}}\chi-\alpha \left(\frac{\xi^2}{2\beta^2}\right)^{-\frac{1}{2\beta^2}}\Gamma\left(1+\frac{1}{2\beta^2},\frac{\xi^2}{2\beta^2}\right)\right]\bigg{\}}.\label{chic}
\end{align}
Using the asymptotic expansion of $\Gamma[a,z]$ for $z \gg 1$, we obtain
\begin{align}
\chi_c=&\text{e}^{\frac{\xi_c^2}{2\beta^2}}\bigg{\{}\alpha \left(\frac{\xi_c^2}{2\beta^2}\right)^{-\frac{1}{2\beta^2}}\Gamma\left(1+\frac{1}{2\beta^2},\frac{\xi_c^2}{2\beta^2}\right)\nonumber\\
&+\left(\frac{\xi}{\xi_c}\right)^{\frac{1}{\beta^2}}\text{e}^{-\frac{\xi^2}{2\beta^2}}\left[\chi-\alpha +O\left(\left(\frac{\xi^2}{2\beta^2}\right)^{-2}\right)\right]\bigg{\}}.
\end{align}
Therefore, the complete expression for the $e$-folding number is
\begin{align}
     N_2-N&=\frac{1}{4}\left(\frac{\varphi_c}{M_p}\right)^2(\xi^2-\xi_c^2)+\frac{1}{\eta_\varphi}\ln\left(\frac{\xi}{\xi_c}\right)+\frac{1}{\eta_\theta}\ln\left[\frac{\chi_c(\xi,\chi,\xi_c)}{\chi_2}\right]+\frac{1}{2\eta_\varphi}\xi_c^2+\cdots.
\end{align}
Here, terms suppressed exponentially have been omitted. Consequently, the $N_\theta$ term, which influences the growth of the power spectrum, is given by
\begin{align}
    N_\theta=\frac{1}{\eta_{\theta}\varphi_c\alpha ({2e\beta^2})^{\frac{1}{2\beta^2}}\Gamma(1+\frac{1}{2\beta^2},\frac{1}{2\beta^2})}\left(\frac{\xi}{\xi_c}\right)^{\frac{1}{\beta^2}}\text{e}^{-\frac{\xi^2-\xi_c^2}{2\beta^2}}\propto \exp\left(-\frac{\xi^2}{2\beta^2}\right).
\end{align}
This shows that the contribution from the waterfall direction is exponentially suppressed away from the critical point and becomes important only when $\xi$ approaches $\xi_c$.
%%%%%%%%%%%%%%%%%%%%%%%%%%%%%%%%%%%%%%%%%%%%%%%%%%%%%% %%%%%%%%%%%%%%%%%%%%%%%%%%%%%%%%%%%%%%%%%%%%%%%%%%%%%% %%%%%%%%%%%%%%%%%%%%%%%%%%%%%%%%%%%%%%%%%%%%%%%%%%%%%% 
\subsection{Enhancement mechanism of curvature perturbation}\label{mecha}
\begin{figure}[htbp]
\centering
    \includegraphics[width=\textwidth]{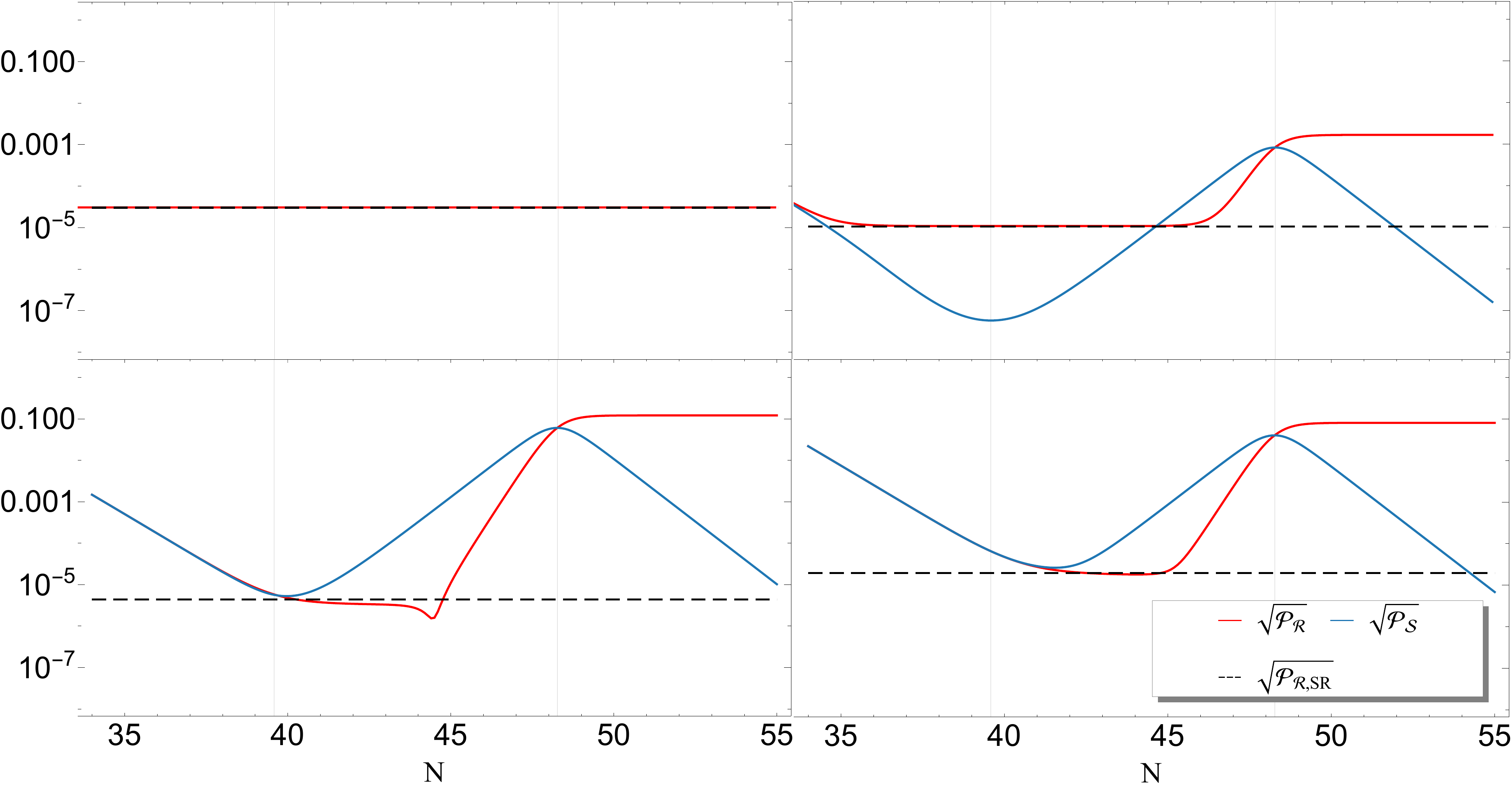}
  \caption{The four different wavenumbers correspond to perturbations exiting the horizon far before, close to, at, and after the critical point, respectively. SR approximation uses $\sqrt{\mathcal{P}_{\mathcal{R,\rm SR}}(k)}=\sqrt{H^2/(8\pi^2\epsilon M_p^2)}$ 
  at horizon exit to approximate the curvature perturbation power spectrum. The vertical grid lines, from left to right, represent the critical point and the field-space turning point.
  }
    \label{difk}
\end{figure}

To study the enhancement mechanism of curvature perturbation, we select some Fourier modes and show their evolution in \autoref{difk}.

We define $N_c$ as the beginning of Stage-2 (the time when the effective mass of the $\theta$ field becomes zero, so called critical point) and $N_\mathrm{eq}$ as the time when the absolute values of the field velocities are equal.  It can be clearly seen that for perturbations which exit the horizon far before $N_c$ , the curvature perturbation remains conserved after horizon exit. Although the isocurvature perturbation grows, 
its amplitude is too small to affect the evolution of the curvature perturbation. For perturbations exiting the horizon shortly before $N_c$, the curvature perturbation exhibits enhancement near the $N_\mathrm{eq}$, 
as the isocurvature perturbation grows and becomes large enough to influence the curvature perturbation. Subsequently, the isocurvature perturbation decays, and the curvature perturbation remains conserved thereafter. For perturbations exiting the horizon after the $N_c$, although the curvature perturbation still experiences some growth, the duration is reduced, leading to an enhancement smaller than the peak value. Next, we will analyze this phenomenon from two perspectives: linear perturbation theory and the phase-space formalism, and the two can be cross-checked for consistency.

\subsubsection{Linear perturbation theory}
In linear perturbation theory, one typically starts with two scalar fields ${\phi^I}$, where $I = \varphi, \theta$. Using the field velocities $\pi^I = \partial\phi^I/\partial N$, the adiabatic direction $\hat{\sigma}^I = \pi^I / \sqrt{\pi^2}$ and the isocurvature direction $\hat{s}^J = \epsilon^{IJ} \hat{\sigma}_I$ are defined in field-space, where $\epsilon^{IJ}$ is the totally antisymmetric tensor. It is straightforward to verify that these are two orthogonal unit vectors. The time evolution of these unit vectors is captured by the turn rate pseudovector and pseudoscalar \cite{Lorenzoni:2025gni,McDonough:2020gmn}
\begin{align}
     \omega^I=\mathcal{D}_t\hat{\sigma}^I,\quad \omega=\epsilon_{IJ}\hat{\sigma}^I\omega^J=\hat{s}_J\omega^J.
\end{align}
Similarly, the scalar field perturbations on spatial flat hypersurface can be projected onto these two directions, decomposing them into the adiabatic perturbation (curvature perturbation) $\mathcal{R}_k$ and the isocurvature perturbation $\mathcal{S}_k$, where the subscript $k$ denotes the Fourier mode. This leads to the evolution equations for these two gauge-invariant perturbations on super-horizon scales \cite{Kaiser:2012ak,Achucarro:2016fby,Lorenzoni:2024krn,Fumagalli:2020adf}
\begin{align}\label{eq:R'=S}
     & \mathcal{R}'_k=2\frac{\omega}{H}\mathcal{S}_k,\\
    & {\mathcal{S}}''_k+(3+\epsilon_2-\epsilon)\mathcal{S}'_k+\frac{\mu_{ss
}^2}{H^2}\mathcal{S}_k=0,\label{eomS}\\
&\mu_{ss}^2=\mathcal{V}_{ss}+3\omega^2+\frac{\epsilon_2 H^2}{2}(3-\epsilon+\frac{1}{2}\epsilon_2+\epsilon_3),\\
&\mathcal{V}_{ss}=\hat{s}^I\hat{s}^JV_{IJ},
\end{align}
where $\epsilon_i$ denote the Hubble-flow slow-roll parameters.
On the right hand side of \eqref{eq:R'=S}, it is clearly seen that the product of the isocurvature perturbation $\mathcal{S}_k$ and the turning rate $\omega$ acts as the source term for the evolution of the curvature perturbation, while the isocurvature perturbation sources itself. 
This indicates that the conversion from isocurvature to curvature perturbation occurs only when there is a turn in the field space. Otherwise as the turn rate negligible, the curvature perturbation remains sourceless and conserved. Furthermore, when the isocurvature effective mass-square $\mu_{ss}^2$ is positive, $\mathcal{S}_k$ typically behaves as a decaying mode. 
When $\mu_{ss}^2$ is negative, the isocurvature perturbation experiences tachyonic instability and thus acts as a source leading to the growth of the curvature perturbation.

We compute the evolution of the effective mass of the isocurvature perturbation and the turn rate in different epochs 
 \begin{align}
    \frac{\mu_{ss}^2}{H^2}&=\begin{cases}
        \displaystyle\frac{6}{\beta^2}\left(\frac{\varphi_c}{M_p}\right)^{-2},&\quad N< N_c,\\
        \\
        3\eta_\theta-\lambda_\varphi(3-\lambda_\varphi),&\quad N_c<N<N_\mathrm{eq},\\
        \\
       3\eta_\varphi- \lambda_\theta(3-\lambda_\theta),&\quad N>N_\mathrm{eq}.
    \end{cases}
    \\
    \nonumber\\
    \frac{\omega}{H}&=\begin{cases}
   \sim  0,&\quad N<N_c,\\
   \\
   \displaystyle-(\lambda_\varphi-\lambda_\theta)\frac{\pi_\theta}{\pi_\varphi}\propto\exp[(\lambda_\varphi-\lambda_\theta)N],&\quad N_c<N<N_\mathrm{eq},\\
   \\
     \displaystyle-(\lambda_\varphi-\lambda_\theta)\frac{\pi_\varphi}{\pi_\theta}\propto\exp[-(\lambda_\varphi-\lambda_\theta)N],&\quad N>N_\mathrm{eq}.
    \end{cases}
\end{align}
Therefore, the contribution of the turn rate to the effective mass of the isocurvature perturbation is negligible. However, its dependence on $N$ plays an important role in the conversion process from isocurvature perturbation to curvature perturbation.

Finally, we obtain the evolution of isocurvature perturbation and curvature perturbation
\begin{align}
    \mathcal{S}_k&\propto \begin{cases}
        \displaystyle\exp\left[-\left(\frac{3}{2}-\sqrt{\frac{9}{4}+\frac{6\eta_\theta}{\eta_\varphi}\left(\frac{\varphi_c}{M_p}\right)^{-2}}\right)N\right],&\quad N<N_c,\\
        \\
        \exp\big[(\lambda_\varphi-\lambda_\theta)N\big], &\quad N_c<N<N_\mathrm{eq},\\
        \\
          \exp\big[-(\lambda_\varphi-\lambda_\theta)N\big],&\quad N>N_\mathrm{eq}.\label{SN}
    \end{cases}\\
    \nonumber\\
    \mathcal{R}_k&\propto\begin{cases}
        \mathrm{const},& \quad N\ll N_\mathrm{eq}\quad {\rm and} \quad N\gg N_\mathrm{eq},\\
        \\
     \mathrm{const}+    \exp\left[2(\lambda_\varphi-\lambda_\theta) N\right],&\quad N\sim N_\mathrm{eq}^-,\\
     \\
     \mathrm{const}-    \exp\left[-2(\lambda_\varphi-\lambda_\theta) N\right],&\quad N\sim N_\mathrm{eq}^+.\label{RN}
    \end{cases}
\end{align}

From this, it is evident that the growth of the isocurvature perturbation leads to an enhancement of the curvature perturbation. This originates from the $\theta$ field rolling down a potential with $V_{\theta\theta}<0$ after the beginning of Stage-2 and before the turn of the field-space trajectory. This rolling renders $\mu_{ss}^2$ negative, thus triggers a phase of tachyonic instability. During the turn, the isocurvature perturbation acts as a source, driving the growth of the curvature perturbation. Before the turn, $\mu_{ss}^2$ is basically determined by that of the scalar field $\theta$. Consequently, as the effective mass-square of $\theta$ transitions from positive to negative, $\mu_{ss}^2$ undergoes a similar change, thereby gives rise to a growth of the curvature perturbation.
  \begin{figure}[H]
  \centering
  \begin{minipage}{0.49\textwidth}
    \centering
    \includegraphics[width=\textwidth]{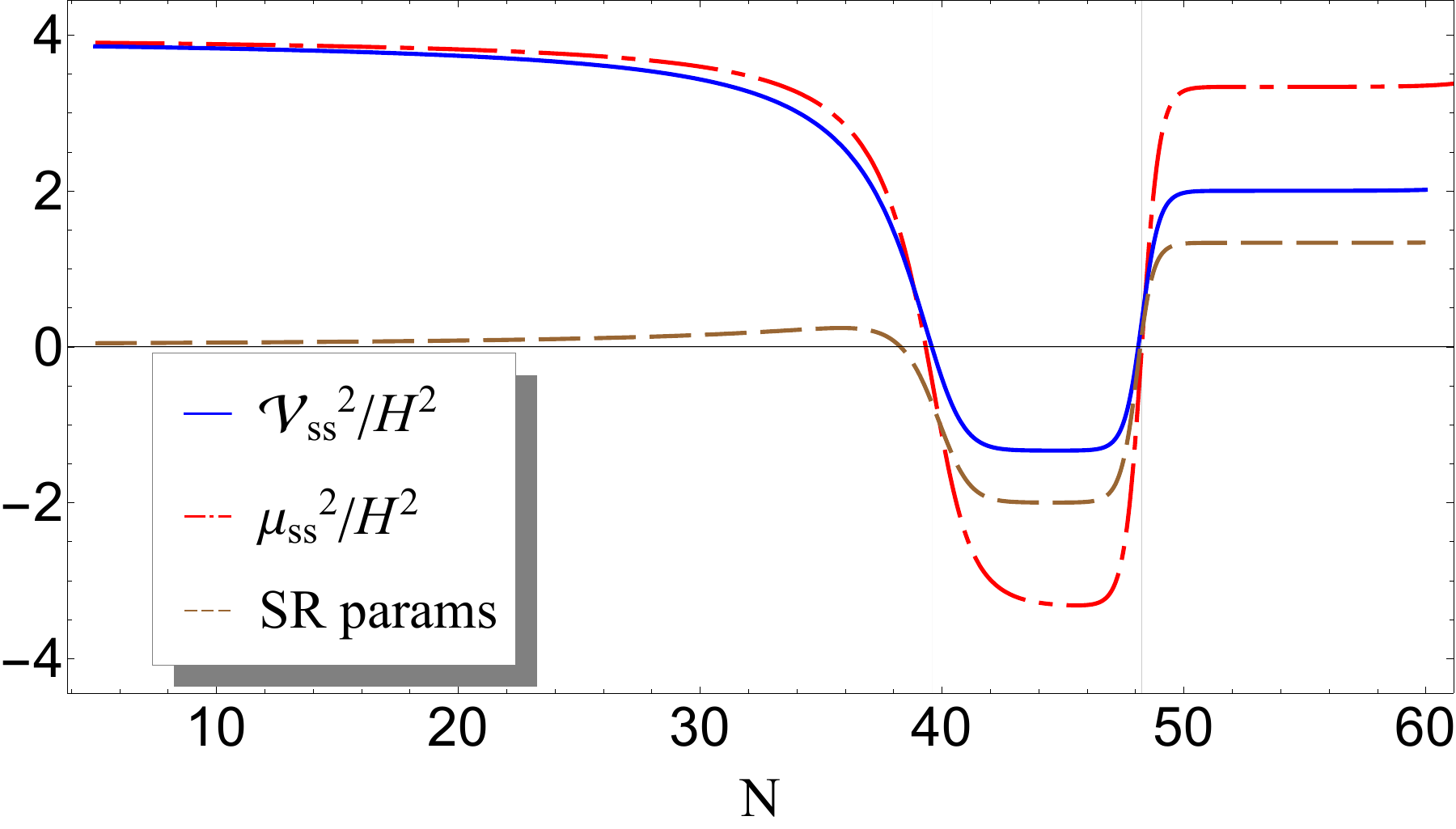} % 第二张图片
  %  \caption{第二张图片的标题}
    \label{fig:image2}
  \end{minipage}
  \begin{minipage}{0.49\textwidth}
    \centering
    \includegraphics[width=\textwidth]{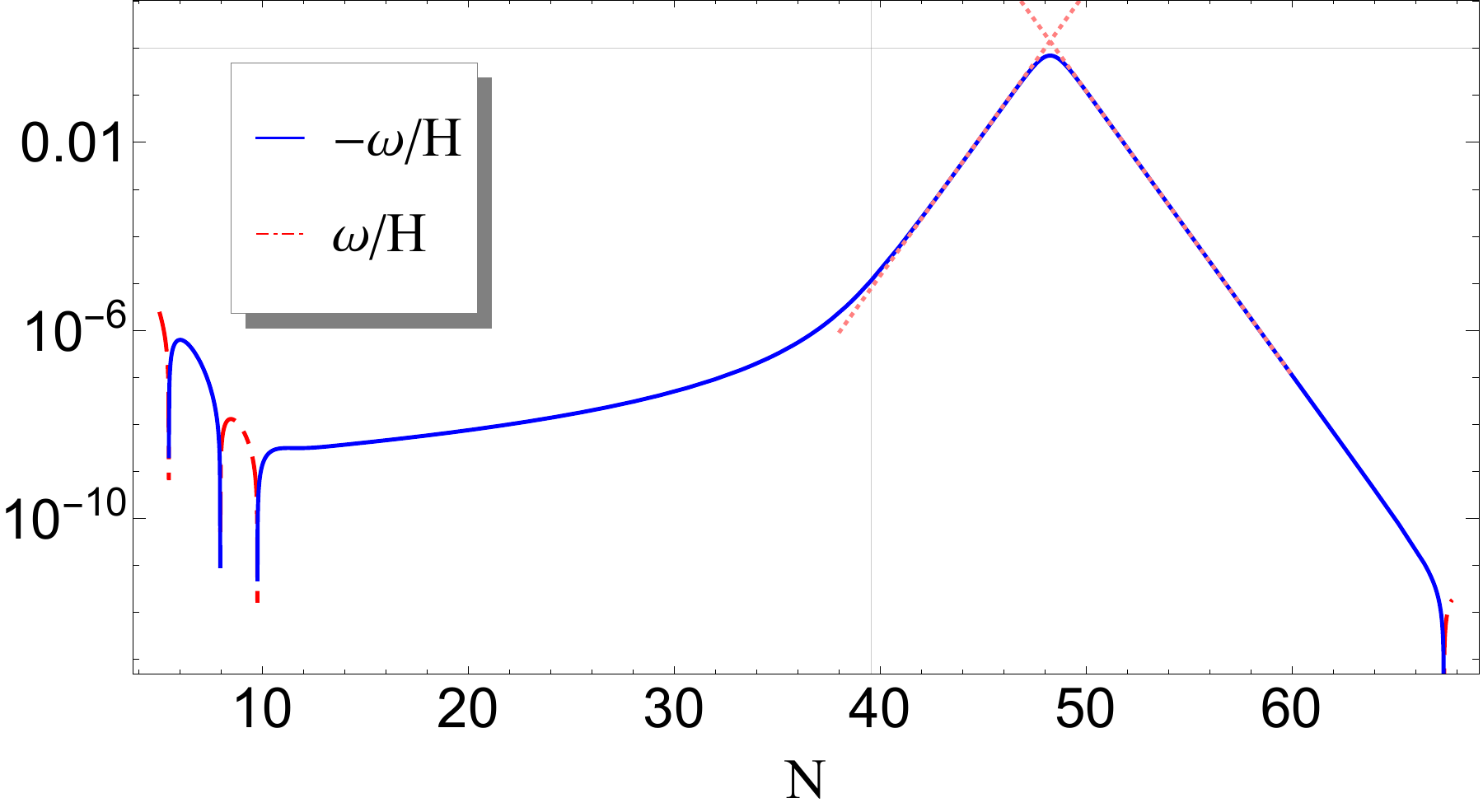} % 第二张图片
  %  \caption{第二张图片的标题}
    \label{fig:image3}
  \end{minipage}
  \caption{The left panel depicts the evolution of the parameters in the effective mass-square term, while the right panel presents the evolution of the turn rate in field-space. The vertical grid lines, from left to right, represent the critical point $N=N_c$ and the field-space turning point $N=N_\mathrm{eq}$ .
  }
  \end{figure}
  
It is crucial to note that an increase in the isocurvature perturbation does not lead to a simultaneous increase in the curvature perturbation. The growing mode becomes dominant only when the isocurvature perturbation is comparable to that of the curvature perturbation, \textit{i.e.}, when $H\mathcal{R}_k\sim\omega \mathcal{S}_k$. During the whole process, the value of the second slow-roll parameter $\epsilon_2$ is always maintained above $-3$, which avoids the emergence of growing modes \cite{Leach:2001zf}. As a result, the enhancement mechanism of the curvature perturbations here is different from the single-field ultra-slow-roll inflation. Instead, the shape of the potential is crucial for the growth of curvature perturbations. However, the estimation error of  the effective mass-square and turn rate
for $N < N_c$ is relatively large because the dependence of the parameters on $N$ during this period is quite complicated and cannot be accurately estimated by a constant value.
\begin{figure}[H]
    \centering
    \includegraphics[width=0.8\textwidth]{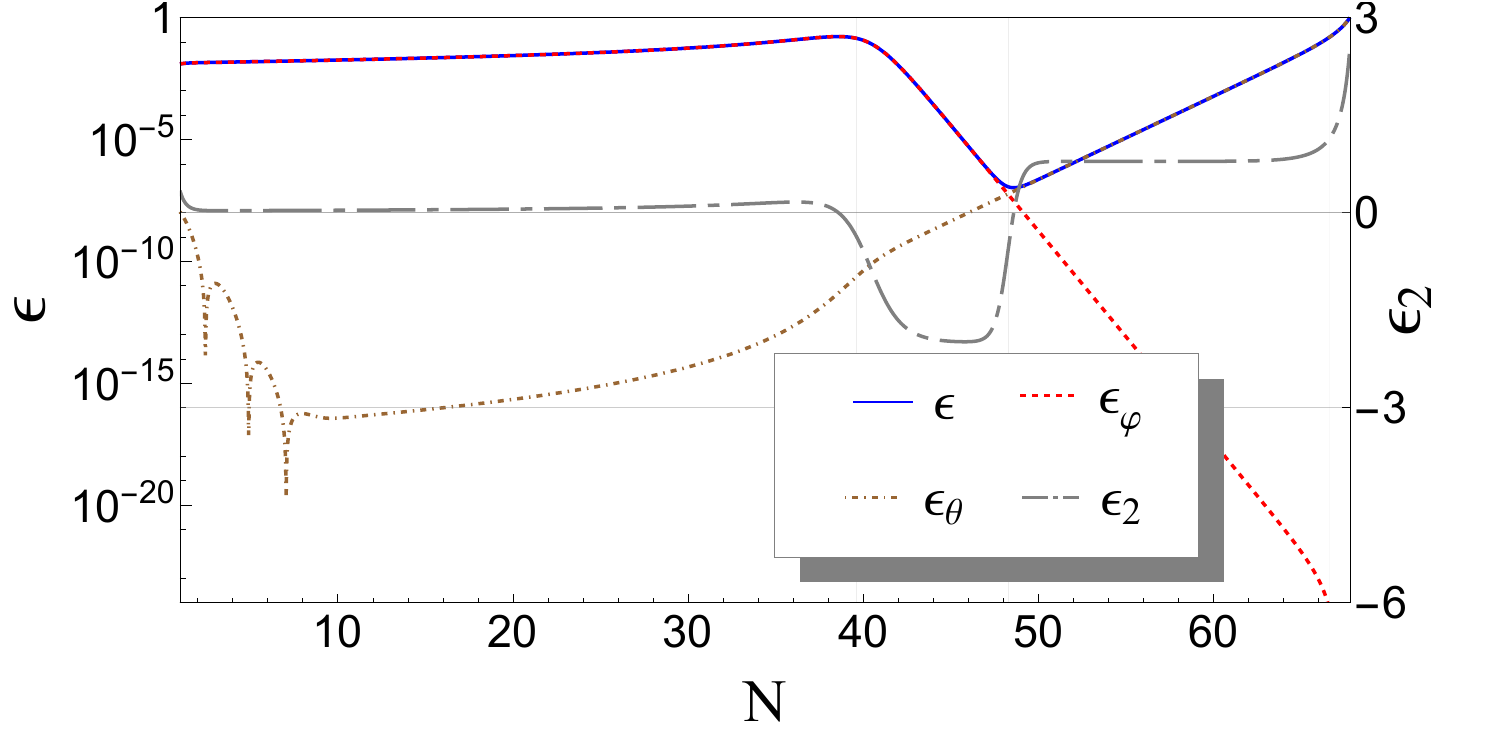} % 第二张图片
    \caption{This figure depicts the evolution of slow roll parameters. The vertical grid lines, from left to right, represent the critical point, the field-space turning point and the end of inflation. The parallel grid line represents the condition $\epsilon_2 = -3$.}
    \label{slrp}
\end{figure}
\begin{figure}[H]
    \centering
    \includegraphics[width=0.8\textwidth]{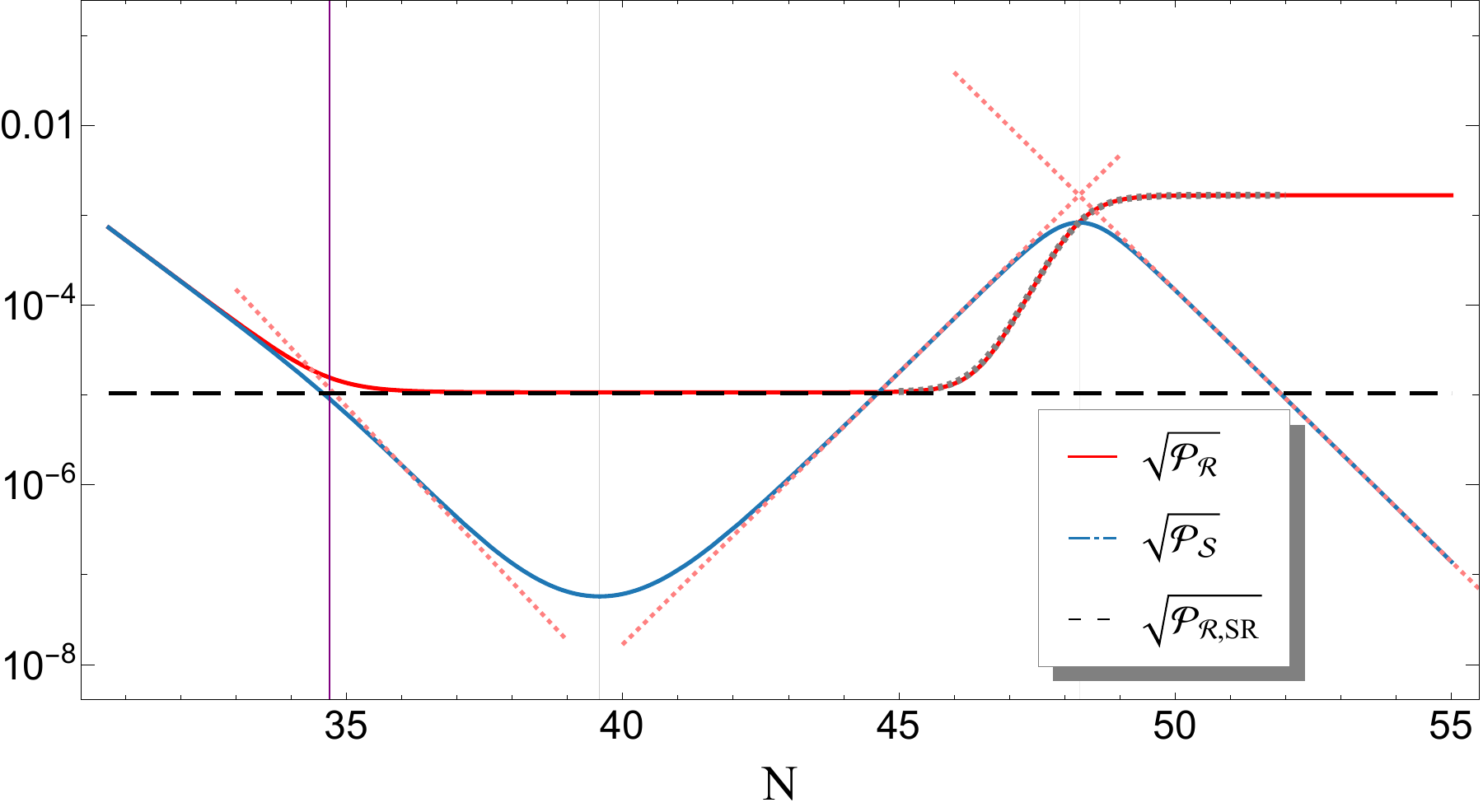} % 第二张图片
  %  \caption{第二张图片的标题}
    \label{k107ev}
  \caption{This figure shows the evolution of curvature perturbation and the dashed line represents the behavior as given in \eqref{SN} and \eqref{RN}. The vertical grid lines, from left to right, represent the critical point and the field-space turning point, while the vertical solid line indicates the horizon-exit time (purple).
  }
\end{figure}
  
From the above discussion, we can also discuss the $k$-dependence of the power spectrum $\mathcal{P}_\mathcal{R}(k)$. When $k > k_c$, the spectral shape of $\mathcal{P}_\mathcal{R}(k)$ is determined by the effective single-field inflation of $\theta$ field with a potential $V(\theta) = V_0 \left(1 + \frac{1}{2} \eta_\theta (\theta/M_p)^2\right)$, thus the spectral index is $n_s - 1 = 2 \lambda_\theta$. For $k < k_c$, the power spectrum cannot be simply attributed to an effective single-field model. Specifically, we can study this issue by examining the evolution of the isocurvature perturbation: for perturbations that enter the horizon some time before the critical point, their evolution after horizon exit and before reaching the critical point is approximately described by the first line of
When $\beta^2(\varphi_c/M_p)^2 < 8/3$
(\textit{i.e.} the isocurvature perturbation has a large effective mass), the characteristic roots of the isocurvature perturbation equation of motion \eqref{eomS} are complex, 
leading to a damped oscillation with a decaying amplitude $\propto a^{-3/2}$. Furthermore, during the period after the critical point but before the turn, the amplification factor of the isocurvature perturbation remains constant.

The isocurvature perturbation acts as a source for the curvature perturbation, resulting in a curvature perturbation $\mathcal{R}_k \sim \exp\left[3(N_k - N_c)/2\right]$ for perturbations that enter the horizon before the critical point. Consequently, the power spectrum of the curvature perturbation in this regime typically exhibits a $k^3$ growth, which originates in the large effective mass $\mu_{ss}$ of the isocurvature perturbation. We see that if the mass $\mu_{ss}$ is not large, the exponents in the first line of \eqref{SN} could be real, resulting a different $k$-dependence in $\mathcal{P_R}(k)$, which we will not discuss in this paper for simplicity.
Therefore, the power spectrum around peak can be approximated by the following analytic expression
\begin{align}
\mathcal{P}_\mathcal{R}(k)=A_\mathcal{R}\left(\frac{k}{k_c}\right)^3\left[1+B\left(\frac{k}{k_c}\right)^{\frac{1}{\Delta}}\right]^{\Delta(-3+2\lambda_\theta)}.\label{powerP}
\end{align}
Here, $A , B$ and $\Delta$ are coefficients that cannot be determined by the simple arguments above. Interestingly, the ultraviolet spectral tilt $2\lambda_\theta$ is twice the attractor characteristic root of the waterfall field, which, as we will see later, can be observed directly from the induced GW energy spectrum.

\subsubsection{Phase-space formalism}
Starting from the expressions for the curvature and isocurvature perturbations \eqref{RDNDM} and \eqref{SDNDM}, we use the field-space coordinate $M$ and $N$ to study the evolution of perturbations. When $N > N_c$, the coordinates are chosen as $(N, M = \xi_f)$. The hypersurface of the final state at each moment is a surface of constant $\chi$, thus
\begin{align}
    \mathcal{R}_c(N)
    % &=\Delta N{-}\left(\frac{\pi_\varphi\frac{\partial\varphi}{\partial\xi_f}+\pi_\theta\frac{\partial\theta}{\partial\xi_f}}{\pi^2}\right)_N\Delta\xi_f\nonumber\\
    &=\Delta N-\left(\frac{\pi_\varphi}{\pi^2}\right)_N\varphi_c\Delta\xi_f,\qquad
    \mathcal{S}(N)
    % &=\left(\frac{-\pi_\theta\frac{\partial\varphi}{\partial\xi_f}+\pi_\varphi\frac{\partial\theta}{\partial\xi_f}}{\pi^2}\right)_N\Delta\xi_f\nonumber\\
    =-\left(\frac{\pi_\theta}{\pi^2}\right)_N\varphi_c\Delta\xi_f.
\end{align}
When $N < N_c$, the coordinates are chosen as $(N, M = \chi_f)$, with the final-state hypersurface being a surface of constant $\xi$,
\begin{align}
    \mathcal{R}_c(N)
    % &=\Delta N-\left(\frac{\pi_\varphi\frac{\partial\varphi}{\partial\chi_f}+\pi_\theta\frac{\partial\theta}{\partial\chi_f}}{\pi^2}\right)_N\Delta\chi_f\nonumber\\
    &=\Delta N-\left(\frac{\pi_\theta}{\pi^2}\right)_N\varphi_c\Delta\chi_f,\qquad
    \mathcal{S}(N)
    % &=\left(\frac{-\pi_\theta\frac{\partial\varphi}{\partial\chi_f}+\pi_\varphi\frac{\partial\theta}{\partial\chi_f}}{\pi^2}\right)_N\Delta\chi_f\nonumber\\
    =\left(\frac{\pi_\varphi}{\pi^2}\right)_N\varphi_c\Delta\chi_f.
\end{align}
Next we consider the most complex scenario, where the perturbations exit the horizon before Stage-2.

Firstly we calculate $\Delta M\equiv\Delta\chi_f$ when $N<N_c$. Here $\chi_f$ which means the final value of $\chi$, which is given by \eqref{chic}
\begin{align}
  \chi_f&\approx\text{e}^{\frac{\xi_f^2}{2\beta^2}}\bigg{[}\alpha \left(\frac{\xi_f^2}{2\beta^2}\right)^{-\frac{1}{2\beta^2}}\Gamma\left(1+\frac{1}{2\beta^2},\frac{\xi_f^2}{2\beta^2}\right)+\left(\frac{\xi}{\xi_f}\right)^{\frac{1}{\beta^2}}\text{e}^{-\frac{\xi^2}{2\beta^2}}\left(\chi-\alpha \right)\bigg{]}.
\end{align}
Calculating the linear-order coefficients of $\Delta M$ are
\begin{align}
    M_\xi=\frac{\xi^{-1}-\xi}{\beta^2}\left(\frac{\xi}{\xi_f}\right)^{\frac{1}{\beta^2}}\text{e}^{\frac{\xi_f^2-\xi^2}{2\beta^2}}(\chi-\alpha),\qquad
    M_\chi=\left(\frac{\xi}{\xi_f}\right)^{\frac{1}{\beta^2}}\text{e}^{\frac{\xi_f^2-\xi^2}{2\beta^2}}.\label{Mxichi}
\end{align}
Thus, the expression for the isocurvature perturbation $\mathcal{S}$ at this stage is 
\begin{align}
    \mathcal{S}(N)
   \approx-\left(\frac{1}{2}\xi\left(\frac{\varphi_c}{M_p}\right)^2+\frac{1}{\eta_\varphi\xi}\right){\xi}^{-\frac{1}{\beta^2}}\text{e}^{\frac{\xi^2}{2\beta^2}}\left(\frac{\xi^{-1}-\xi}{\beta^2}(\chi-\alpha)\xi^{\frac{1}{\beta^2}}\text{e}^{-\frac{\xi^2}{2\beta^2}}\delta\xi+\xi^{\frac{1}{\beta^2}}\text{e}^{-\frac{\xi^2}{2\beta^2}}\delta\chi\right)_{N_0}.
\end{align}

For $N>N_c$, the perturbation of $\xi_f$ mainly comes from Stage-2, hence we have \eqref{xi2}:
\begin{align}
    \xi_f^2&=-\frac{1}{s}W_0[Y(\xi_c,\chi_c,\chi_f)],\\
    Y&=-s\exp\left[-s\xi_c^2+2\ln\xi_c+2\beta_e^2\ln\left(\frac{\chi_c}{\chi_f}\right)\right],
\end{align}
where $\chi_c = \chi_c(\xi, \chi, \xi_c)$. We obtain the linear-order coefficients of $\Delta M$
\begin{align}
    M_\xi&=\frac{\xi_f}{1-\xi_f^2}\frac{\beta_e^2}{\chi_c}\frac{\partial \chi_c}{\partial\xi},\\
    M_\chi&=\frac{\xi_f}{1-\xi_f^2}\frac{\beta_e^2}{\chi_c}\frac{\partial \chi_c}{\partial\chi}.
\end{align}
The expression for the isocurvature perturbation $\mathcal{S}$ is
\begin{align}
    \mathcal{S}(N)
    % &=-\left(\frac{\pi_\theta}{\pi^2}\right)_N\varphi_c\Delta\xi_f
    &\approx-\frac{\pi_\theta}{\pi^2}\varphi\left(\frac{\beta_e^2}{\chi_c}\frac{\partial \chi_c}{\partial\xi}\delta\xi+\frac{\beta_e^2}{\chi_c}\frac{\partial \chi_c} {\partial\chi}\delta\chi\right)_{N_0} \nonumber\\
  &  \propto\frac{\pi_\theta\pi_\varphi}{\pi^2_\theta+\pi^2_\varphi}=\begin{cases}
        \exp[(\lambda_\varphi-\lambda_\theta)N],&\quad N<N_\mathrm{eq};\\
        \\
        \exp[-(\lambda_\varphi-\lambda_\theta)N],&\quad N>N_\mathrm{eq}.
    \end{cases}\label{isoc} 
\end{align}

Then we calculate $\Delta N$. Since the curvature perturbation only grows during a period around $N \sim N_\mathrm{eq}$ , we primarily calculate the evolution of $\Delta N$ around $N \sim N_\mathrm{eq}$ for modes that exit the horizon before Stage-2. Let the number of e-folds on the final-state hypersurface be $N_f$. 
We have
\begin{align}
   N_f-N_0&=\frac{1}{\lambda_\theta}\ln\frac{\chi_c}{\chi_f}+\frac{\xi_c^2}{2\lambda_\varphi}+\frac{W_0[Y(\xi_c,\chi_c,\chi_f)]}{2\lambda_\varphi}+\left(\frac{\varphi_c}{2M_p}\right)^2(\xi^2-\xi_c^2)+\frac{1}{\eta_\varphi}\ln\frac{\xi}{\xi_c},\\
    N_\xi&=-\left(-\frac{1}{\lambda_\theta\chi_c}+\frac{1}{\lambda_\varphi}\frac{\xi_f^2}{1-\xi_f^2}\frac{\beta_e^2}{\chi_c}\right)\frac{\partial\chi_c}{\partial\xi}+\frac{1}{2}\left(\frac{\varphi_c}{M_p}\right)^2\xi+\frac{1}{\eta_\varphi\xi},\\
    N_\chi&=-\left(-\frac{1}{\lambda_\theta\chi_c}+\frac{1}{\lambda_\varphi}\frac{\xi_f^2}{1-\xi_f^2}\frac{\beta_e^2}{\chi_c}\right)\frac{\partial\chi_c}{\partial\chi},\\\nonumber
    \mathcal{R}_c(N) 
    &=\left[\left(-\left(-\frac{1}{\lambda_\theta\chi_c}+\frac{1}{\lambda_\varphi}\frac{\xi_f^2}{1-\xi_f^2}\frac{\beta_e^2}{\chi_c}\right)\frac{\partial\chi_c}{\partial\xi}+\frac{1}{2}\left(\frac{\varphi_c}{M_p}\right)^2\xi+\frac{1}{\eta_\varphi\xi}\right)\delta\xi\right.\\\nonumber
    &\quad\left.-\left(-\frac{1}{\lambda_\theta\chi_c}+\frac{1}{\lambda_\varphi}\frac{\xi_f^2}{1-\xi_f^2}\frac{\beta_e^2}{\chi_c}\right)\frac{\partial\chi_c}{\partial\chi}\delta\chi\right]_{N_0}-\frac{\pi_\varphi}{\pi^2}\varphi\left(\frac{\beta_e^2}{\chi_c}\frac{\partial \chi_c}{\partial\xi}\delta\xi+\frac{\beta_e^2}{\chi_c}\frac{\partial \chi_c}{\partial\chi}\delta\chi\right)_{N_0}\\
    &=\Delta N+\frac{\pi_\varphi}{\pi_\theta}\mathcal{S}(N)\equiv\mathcal{R}_{\delta\theta=0}(N)+\frac{\pi_\varphi}{\pi_\theta}\mathcal{S}(N).\label{cur}
\end{align}
The expression on the RHS of the final equality can also be directly obtained from equation \eqref{RST}. It can be found that the $\Delta N$ part of the curvature perturbation depends solely on the values of the scalar fields when they cross the horizon (except some decay modes depending on $N$). The growth of the curvature perturbation comes from the $\Delta M$ part, while there will be no growth of the curvature perturbation if only $\Delta N$ is considered. The $N$-dependence of $\mathcal{R}$ is
\begin{align}
    \mathcal{R}_c(N)\propto \mathrm{const}-\frac{\pi_\varphi^2}{\pi_\varphi^2+\pi_\theta^2}&=\mathcal{R}_k-\frac{\mathcal{R}_k-\mathcal{R}_{\rm SR}}{1+\exp[2(\lambda_\varphi-\lambda_\theta)(N-N_\mathrm{eq})]},\label{RRN}
\end{align}
where the $\mathcal{R}_k$ is the value of comoving curvature perturbation at the end of inflation and $\mathcal{R}_{\rm SR}$ is given by the slow-roll approximation at horizon exit. 

\begin{figure}[htbp]
\centering
    \includegraphics[width=0.8\textwidth]{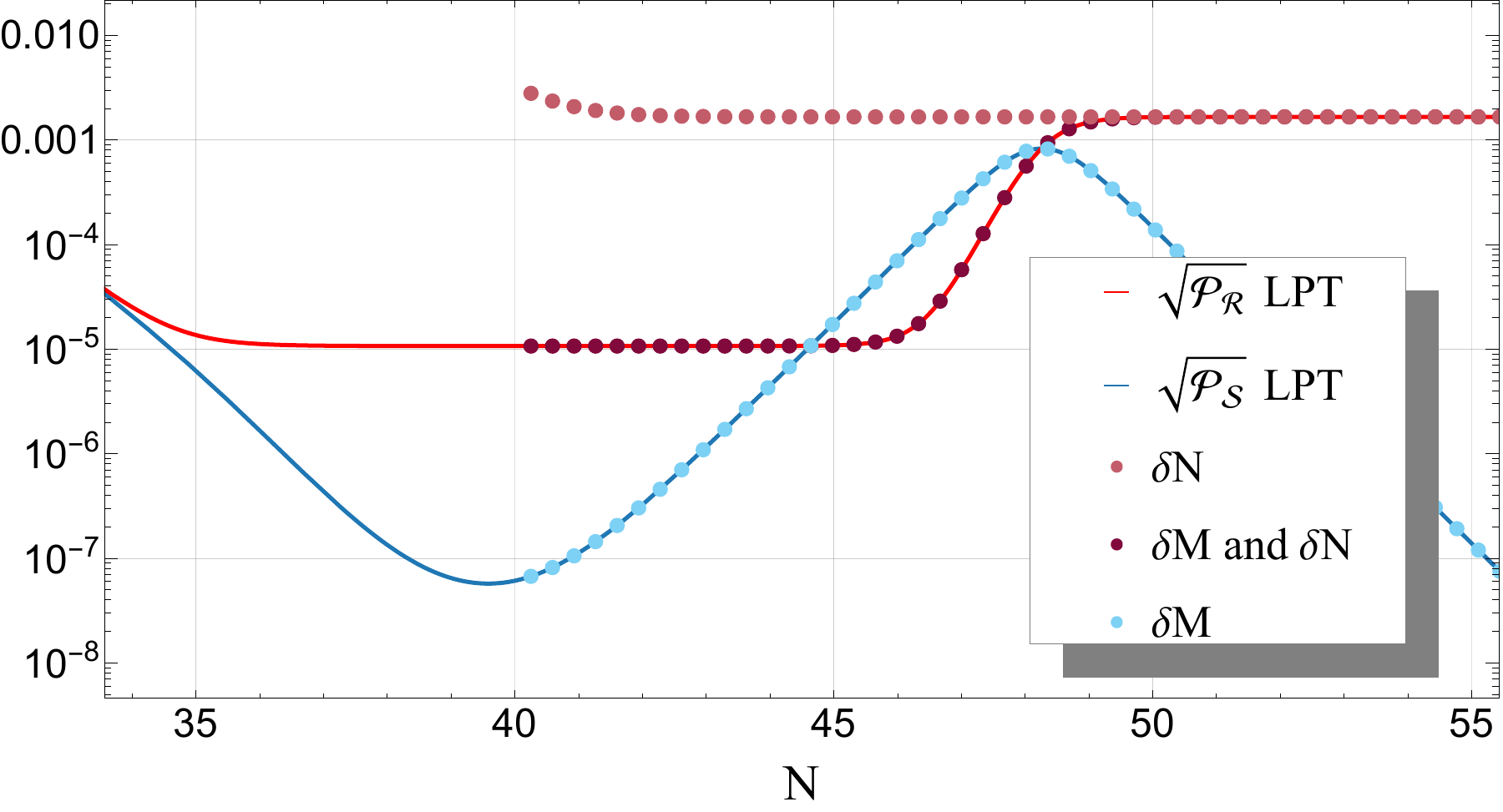}
  \caption{This figure demonstrates the evolution of the curvature perturbation and the isocurvature perturbation for modes that exit the horizon near the critical point and the impact of including the term $\Delta M$ on the evolution of the curvature perturbation, compared to the case where only $\Delta N\,(\mathcal{R}_{\delta\theta =0})$ is considered. All points represent the numerical results from the phase-space formalism with the final hypersurface chosen as the uniform-$\theta$ surface, while the solid lines are from linear perturbation theory. The topmost points correspond to the case where only $\Delta N$ is taken into account.
  }
    \label{dMdN}
\end{figure}

To obtain the above relation, we need to analyze the expression for the comoving curvature perturbation \eqref{cur}. It can be found that the dominant term in this expression during this stage is $\partial \chi_c / \partial \chi$, because the change of $\xi$ has almost no effect on $\chi_c$ (as can be seen from \eqref{Mxichi}). Thus, we have
\begin{align}
    \mathcal{R}_c(N)\approx-\left(-\frac{1}{\lambda_\theta\chi_c}+\frac{\pi_\varphi}{\pi^2}\varphi\frac{\beta_e^2}{\chi_c}\right)_N\frac{\partial \chi_c}{\partial\chi}\delta\chi
\end{align}
Only terms in the brackets depend on $N$, which can be easily found in \eqref{RRN}. The other part contributes a N-independent factor. The discussion above is completely independent of linear perturbation theory, which displays the evolution of the perturbations based solely on the background evolution of the model. This precisely illustrates one of the key advantages of $\delta N$ formalism.

%%%%%%%%%%%%%%%%%%%%%%%%%%%%%%%%%%%%%%%%%%%%%%%%%%%%%%%%%%%%%
%%%%%%%%%%%%%%%%%%%%%%%%%%%%%%%%%%%%%%%%%%%%%%%%%%%%%%%%%%%%%
%%%%%%%%%%%%%%%%%%%%%%%%%%%%%%%%%%%%%%%%%%%%%%%%%%%%%%%%%%%%%
%\subsection{Phenomenology}\label{GWPBH}
\subsection{Non-Gaussianity}\label{sec:NG}
The formation of PBHs exhibits a strong dependence on the probability distribution function of the curvature perturbation. From \eqref{N} we can obtain the non-linear relationship between the curvature perturbation and the field perturbations
\begin{align}
    \mathcal{R}_c
    &=\frac{1}{2\lambda_\varphi}(2\xi\delta\xi+\delta\xi^2)+\frac{1}{\lambda_\theta}\ln\left(1+\frac{\delta\chi}{\chi}\right)+\frac{1}{2\lambda_\varphi}(W_0[Y(\xi+\delta\xi,\chi+\delta\chi,\chi_f)]-W_0[Y(\xi,\chi,\chi_f)]).\label{rnl}
\end{align}
We can derive the probability distribution of the curvature perturbation at the peak
\begin{align}
    \mathbb{P}[\mathcal{R}_c]&=\iint\text{d}\delta\xi\text{d}\delta\chi\, \mathbb{P}[\delta\xi,\delta\chi]\delta\Big(\mathcal{R}_c-\mathcal{R}_c(\xi_c,\chi_c,\delta\xi,\delta\chi)\Big) \nonumber\\
    &\approx\frac{|\lambda_\theta\chi_c|\text{e}^{\lambda_\theta\mathcal{R}_c}}{\sqrt{2\pi}\sigma_{\delta\chi}}\exp\left[-\frac{\chi_c^2(\text{e}^{\lambda_\theta\mathcal{R}_c}-1)^2}{2\sigma_{\delta\chi}^2}\right].\label{PDF}
 \end{align}
 Here, the approximation holds around the peak of the power spectrum, which indicates that the curvature perturbation  \eqref{rnl} is dominated by the logarithmic term
 \begin{align}
    \mathcal{R}_c=\frac{1}{\lambda_\theta}\ln\left(1+\lambda_\theta\mathcal{R}_g\right)
\end{align}
with a nominal nonlinear parameter $f_\mathrm{NL}=-\frac56\lambda_\theta$  which is the leading order of \eqref{fnl}.  This is because near the peak of the power spectrum we have $\chi/\chi_2 \ll 1$, so that higher-order terms are significantly suppressed.

\begin{figure}[htbp]
\centering
    \includegraphics[width=0.8\textwidth]{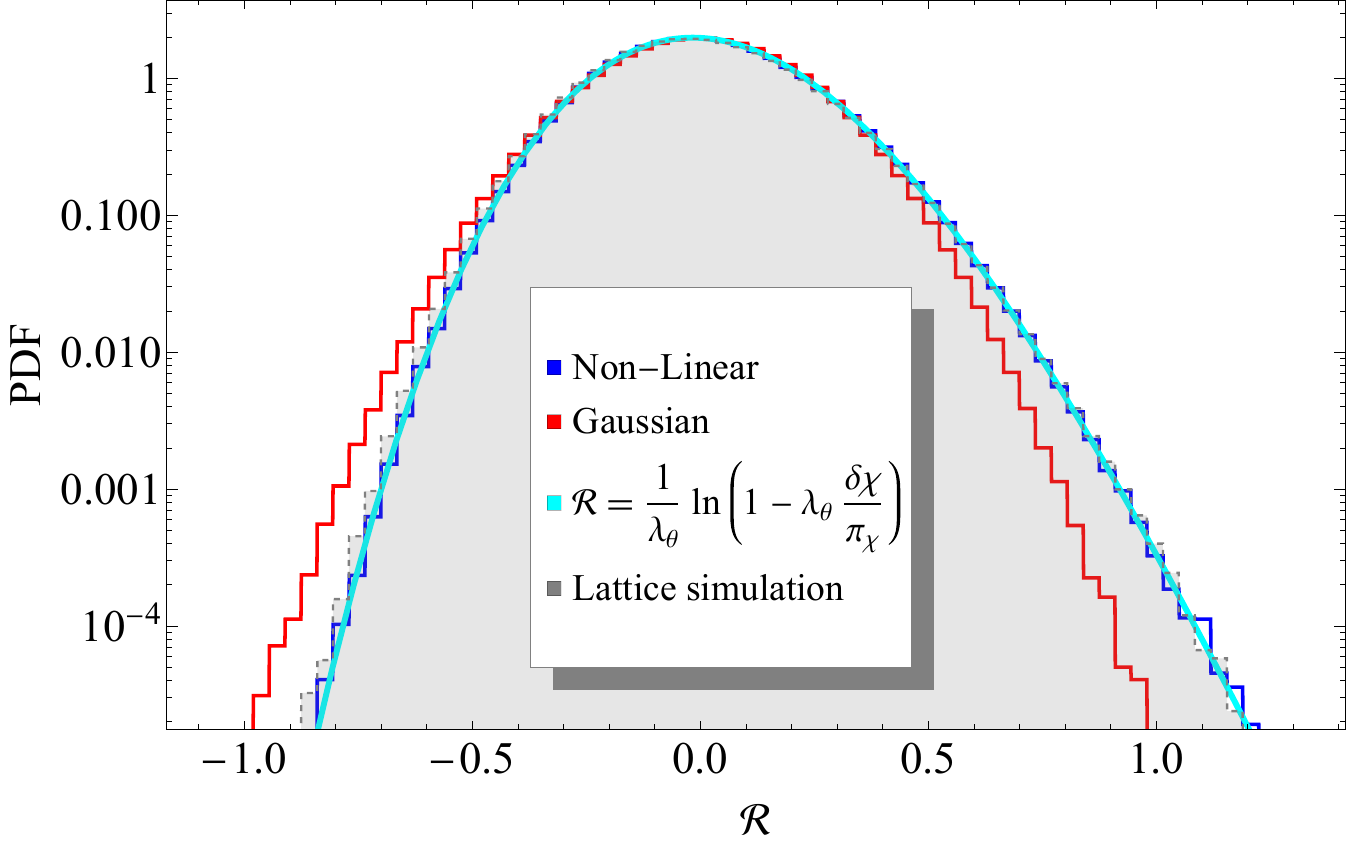}
  \caption{PDF of the comoving curvature perturbation $\mathcal{R}$ on the scale $k_c^{-1}$. 
The analytical result (cyan) is compared with non-linear formula \eqref{rnl} (blue histogram), lattice simulations (gray histogram) and the Gaussian distribution (red histogram).
 The parameters are chosen as follows $\eta_\varphi=2/3,\eta_\theta=-4/9,\beta^2=3/2,\alpha=1/40000,\xi_c=1$ which give $\lambda_\theta\approx-0.39$.} 
    \label{fig:pdf}
\end{figure}

The red and blue histograms in \autoref{fig:pdf} are obtained from the same Gaussian realizations of $\delta\chi$ and $\delta\xi$. We estimate their variances in a single-scale approximation from the corresponding dimensionless power spectra at the characteristic wavenumber $k_c$. Specifically, $\mathcal P_{\delta I}(k_c)$ is evaluated at the reference time $N_k$, defined implicitly by
$k_c=0.01\,a(N_k)H(N_k)$, and we take
$\sigma_{\delta I}^2\simeq\mathcal P_{\delta I}(k_c)$ for
$I=\chi,\xi$. Applying the linear and nonlinear mappings to the same Gaussian samples yields the red and blue histograms, respectively, while the cyan curve represents the analytical PDF given by \eqref{PDF}. For comparison, we also evolve the nonlinear two-field dynamics on a $256^3$ lattice with comoving spacing $\Delta x\simeq k_c^{-1}$, and construct the gray histogram from the distribution of the local $\delta N$ evaluated at the end of inflation.

In this model, the primordial non-Gaussianity is almost entirely determined by the waterfall dynamics after the turn. Around the critical point, the effective potential along the waterfall direction can be approximated by an inverted parabola,
\begin{align}
V(\theta)\simeq V_0-\frac12 M^2\theta^2 ,
\end{align}
so that $V_{\theta\theta}<0$ and hence $\eta_\theta\equiv V_{\theta\theta}/(3H^2)<0$. The evolution of the waterfall perturbation is governed by the characteristic equation
\begin{align}
\lambda^2-3\lambda+3\eta_\theta=0,
\end{align}
whose smaller root defines the attractor exponent $\lambda_\theta<0$. The attractor solution gives
\begin{align}
N_f-N_c
=
-\frac{1}{\lambda_\theta}
\ln\!\left(\frac{\chi_f}{\chi_c}\right),
\qquad
f_{\rm NL}
=
-\frac56\,\lambda_\theta >0.
\end{align}

Crucially, the sign of the non-Gaussianity is robustly fixed by the fundamental requirement that inflation must gracefully exit. To successfully terminate inflation, the waterfall potential must be concave downward after the trajectory turn, corresponding to $V_{\theta\theta} < 0$ (or equivalently $\lambda_\theta < 0$). This kinematic necessity inherently dictates a strictly positive nonlinear parameter, $f_{\mathrm{NL}} > 0$. We emphasize that this positivity is a universal and robust feature of such waterfall dynamics, which systematically guarantees an enhancement in the abundance of primordial black holes.

While the positive sign is an inescapable consequence of the potential's concavity, the magnitude of $f_{\mathrm{NL}}$ is determined by the duration of the post-turn evolution, $\Delta N$, and the initial waterfall field value at the turn, $\chi_c$. For a realistic hybrid inflation model requiring an additional $\Delta N \sim \mathcal{O}(10)$ $e$-folds to end inflation, the parameter evaluates to:
\begin{equation}
f_{\mathrm{NL}} = -\frac{5}{6}\lambda_\theta = \frac{5}{6}\frac{\ln(\chi_f/\chi_c)}{\Delta N} \sim \mathcal{O}\left(\frac{|\ln\alpha|}{10}\right).
\end{equation}
As discussed around Eq.~\eqref{eq:alphaconstraint}, the initial condition $\chi_c$ (parameterized by $\alpha$) is only loosely bounded by avoiding the quantum diffusion domination, which in turn provides an upper bound for $f_\mathrm{NL}$
\begin{equation}
    f_\mathrm{NL}\lesssim 0.1\ln\left(\frac{2\pi|\eta_\theta|\varphi_c}{H}\right).
\end{equation}
Consequently, while $f_{\mathrm{NL}}$ is fundamentally constrained to be positive, its amplitude remains highly tunable and can in principle be large.

%%%%%%%%%%%%%%%%%%%%%%%%%%%%%%%%%%%%%%%%%%%%%%%%%%%%%%%%%%%%%
%%%%%%%%%%%%%%%%%%%%%%%%%%%%%%%%%%%%%%%%%%%%%%%%%%%%%%%%%%%%%
%%%%%%%%%%%%%%%%%%%%%%%%%%%%%%%%%%%%%%%%%%%%%%%%%%%%%%%%%%%%%
\subsection{Primordial Black Hole}\label{sec:PBH}
To calculate the PBH abundance for non-Gaussian curvature perturbation, we use the Press-Schechter-type formalism with the compaction function \cite{Gow:2022jfb,Young:2024jsu,Pi:2024lsu}.  We define the compaction function as
\begin{align}
    \mathcal{C}(t,r,\bm{x})\equiv\frac{2\delta M}{r}=\frac{3(aH)^2}{r}\int_0^r\text{d}\rho\,\rho^2\delta(t,\rho,\bm{x}),
\end{align}
where we set $G=1$, while the density contrast 
on comoving slices
$\delta\equiv\delta\rho/\rho$
is \cite{Harada:2015yda}
% \SP{(In this paper, you did't discuss EoS $w$. So please set $w=1/3$ from the beginning.)}{\green (done)}
\begin{align}
    \delta=-\frac{8}{9}\frac{1}{(aH)^2}\text{e}^{-5\mathcal{R}_c/2}\nabla^2\text{e}^{\mathcal{R}_c/2}.
    %-\frac{4}{3}\cdot\frac{3(1+w)}{5+3w}\frac{1}{(aH)^2}\text{e}^{-5\mathcal{R}_c/2}\nabla^2\text{e}^{\mathcal{R}_c/2}.
\end{align}
Assuming the profile of the curvature perturbation $\mathcal{R}_c$ is spherically symmetric, we can obtain
\begin{align}
   % \mathcal{C}=\mathcal{C}_\ell\left(1-\frac{5+3w}{12(1+w)}\mathcal{C}_\ell\right)\xrightarrow{w=1/3}
   \mathcal{C}=\mathcal{C}_\ell-\frac{3}{8}\mathcal{C}_\ell^2,
\end{align}
and here we define the linear compact function
\begin{align}
   \mathcal{C}_\ell\equiv-\frac{4}{3}r\frac{\partial \mathcal{R}_c}{\partial r}.
\end{align}
Due to the non-linear relation between $\mathcal{R}_c$ and $\mathcal{R}_g$, in order to compute the PDF of $\mathcal{C}_\ell$, we can rewrite it in the following form
\begin{align}
    \mathcal{C}_\ell \equiv \frac{X}{Y}, \quad \text{with} \quad X \equiv -\frac{4}{3}r\mathcal{R}'_{g}(r), \quad Y \equiv 1 + \lambda_\theta\mathcal{R}_{g}. 
\end{align}
The covariance matrix is 
\begin{equation}
\Sigma = 
\begin{pmatrix}
\sigma_X^2 & \rho\sigma_X\sigma_Y \\
\rho\sigma_X\sigma_Y & \sigma_Y^2
\end{pmatrix},
\end{equation}
with
\begin{align}
\sigma_X^2 &= \langle XX \rangle = \left( -\frac{4}{3} \right)^2 \int (kr)^2\mathcal{P}_{\mathcal{R}_{g}}(k) \left( \frac{dj_0}{dz}(kr) \right)^2 W^2(kR_s)\text{d}\ln k, \\
\rho\sigma_X\sigma_Y &= \langle XY \rangle = \left( -\frac{4}{3} \right) \lambda_\theta \int (kr)\mathcal{P}_{\mathcal{R}_{g}}(k) j_0(kr) \frac{dj_0}{dz}(kr)W^2(kR_s) \,\text{d}\ln k, \\
\sigma_Y^2 &= \langle YY \rangle = \lambda_\theta^2 \int \mathcal{P}_{\mathcal{R}_{g}}(k) j_0^2(kr)W^2(kR_s) \,\text{d}\ln k.
\end{align}
where \(\rho\) is the correlation between \(X\) and \(Y\), \(j_0(z) = \sin z/z,z\equiv kr\) is the zeroth order spherical Bessel function. $W(k R_s)\equiv \exp[-(k R_s)^2/2]$  is the window function in Fourier space with smoothing scale $R_s$. The joint PDF of $X$ and $Y$ becomes
\begin{align}
 \mathbb{P}[X,Y
 ]&= \frac{1}{2\pi\sqrt{\det\Sigma}} \exp\left[-\frac{1}{2}(X, Y - 1)\Sigma^{-1} \begin{pmatrix} X \\ Y - 1 \end{pmatrix}\right],
\end{align}
which gives the PDF of $\mathcal{C}_\ell$,
\begin{equation}
 \mathbb{P}[\mathcal{C}_\ell] = \int dXdY \,\mathbb{P}(X,Y)\delta\left(\mathcal{C}_\ell - \frac{X}{Y}\right) = \int dY |Y| \cdot  {P}(X \to \mathcal{C}_\ell Y, Y). 
\end{equation}

A PBH forms once the compaction $\mathcal{C}$ exceeds the threshold $\mathcal{C}_\mathrm{th}$. We restrict our analysis to Type I perturbations, for which $\mathcal{C}_\ell < 4/3$. The abundance of PBHs is then given by
\begin{align}
    \beta_{PBH}(M)=K\frac{(\mathcal{C}_\ell-\frac{3}{8}\mathcal{C}_\ell^2-\mathcal{C}_\mathrm{th})^{\gamma+1}}{\gamma(1-\frac{3}{4}\mathcal{C}_\ell)}\mathbb{P}[\mathcal{C}_\ell].\label{beta}
\end{align}
The linear compaction function $\mathcal{C}_\ell$ is related to PBH mass by a power-law relation given by critical collapse \cite{Choptuik:1992jv,Evans:1994pj,Niemeyer:1997mt}
\begin{align}
    \frac{M(\mathcal{C}_\ell)}{M_H} \sim K (\mathcal{C}- \mathcal{C}_{\text{th}})^\gamma = K \left[ \left(\mathcal{C}_\ell - \frac{3}{8}\mathcal{C}_\ell^2\right) - \mathcal{C}_{ \rm{th}} \right]^\gamma,
\end{align}
where the horizon mass $M_H$ can be normalized e.g. \cite{Kitajima:2021fpq}
\begin{align}
    M_H(k_c)&\approx10^{20}\text{g}\left(\frac{g_\ast}{106.75}\right)^{-1/6}\left(\frac{k_c}{1.56\times10^{13}\text{Mpc}^{-1}}\right)^{-2}.
\end{align}
In our numerical estimate, we adopt a fixed threshold prescription $K=1,\;\mathcal{C}_{\rm th}=0.587,\; \gamma=0.36$ while the characteristic size of the overdense region is fixed as \(r=2.35R_s\) with \(R_s=1/k_c\), following Ref.~\cite{Gow:2022jfb,Pi:2024ert}. The parameters of the curvature perturbation power spectrum are chosen consistently with \eqref{eq:fitPR}. For the ordinary compaction function, \(\mathcal{C}_{\rm th}\) is in principle profile-dependent; here we use \(\mathcal{C}_{\rm th}=0.587\) as an effective threshold within this fixed prescription.

\begin{figure}[htbp]
\centering
    \includegraphics[width=0.8\textwidth]{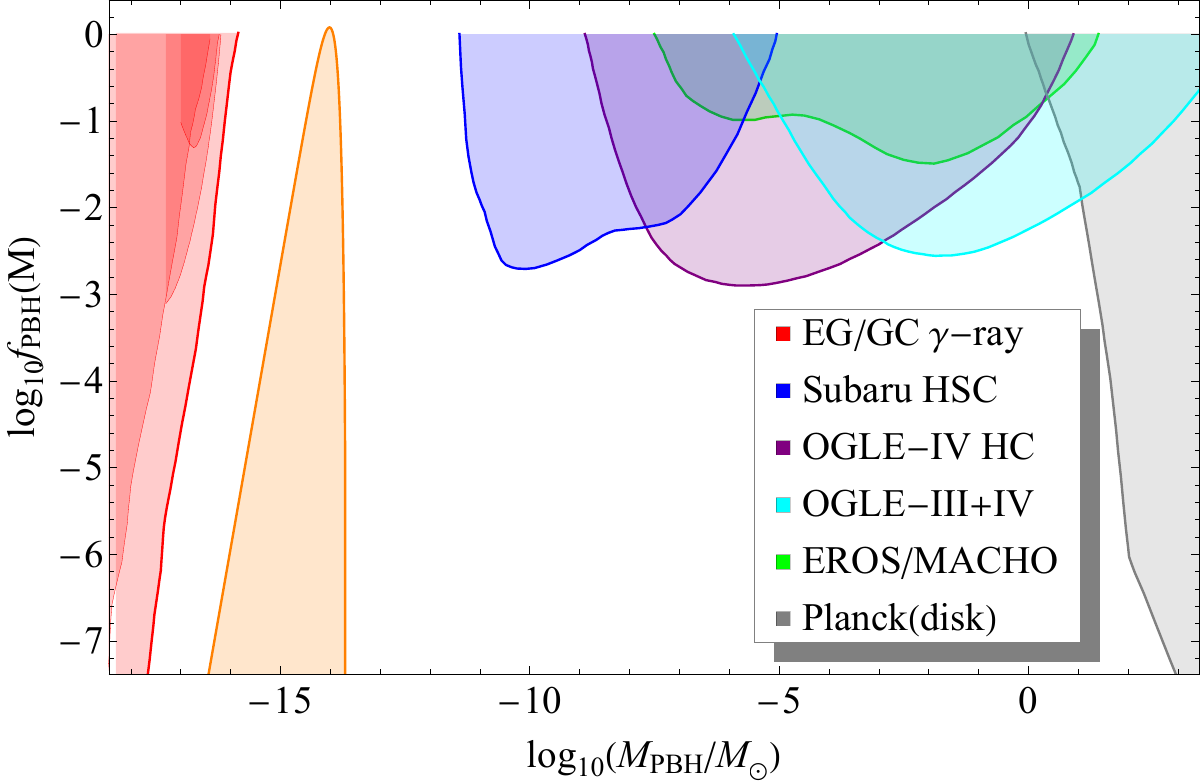}
  \caption{ PBH mass function with current observational constraints. Parameters are the same as in \autoref{GW}. The PBH abundance is normalized such that the total $f_\mathrm{PBH}=1$. 
  }
      \label{pbh}
\end{figure}

Consequently, the PBH mass function is obtained as \cite{Ando:2018qdb}
\begin{align}
    f_\mathrm{PBH}(M)\approx 3.81\times10^8  \left(\frac{g_\ast}{106.75}\right)^{-1/4}\left(\frac{h}{0.67}\right)^{-2}\left(\frac{M}{M_\odot}\right)^{-1/2}\beta_{PBH}(M), 
\end{align}
which is related to the total abundance $f_\mathrm{PBH}$ by
\begin{equation}
    f_\mathrm{PBH}=\int f(M)\mathrm{d}\ln M.
\end{equation}
Currently, observations from different experiments can constrain $f_\mathrm{PBH}$ in different mass ranges. See \autoref{pbh} for details. Notably, the asteroid mass window, \textit{i.e.} $M\sim 10^{17}$--$10^{22}$ g, can not be probed by microlensing or any optical experiments due to the limit of geometric optics and finite-size effect \cite{Sugiyama:2019dgt}. Therefore, asteroid-mass PBHs can account for all dark matter, which can only be explored indirectly by the GWs induced by the enhanced curvature perturbation. Recently, a detection of a few ultrashort timescale microlensing events was reported by Subaru Hyper Suprime-Cam, which suggests that earth-like planetary mass PBH might account for dark matter, which can be crosschecked by multi-messenger GW observations in the future \cite{Domenech:2026nun}. 

%%%%%%%%%%%%%%%%%%%%%%%%%%%%%%%%%%%%%%%%%%%%%%%%%%%%%%%%%%%%%
%%%%%%%%%%%%%%%%%%%%%%%%%%%%%%%%%%%%%%%%%%%%%%%%%%%%%%%%%%%%%
%%%%%%%%%%%%%%%%%%%%%%%%%%%%%%%%%%%%%%%%%%%%%%%%%%%%%%%%%%%%%
\subsection{Induced GWs}\label{sec:IGW}

The amplified curvature perturbation can source the second-order gravitational waves (GWs), which can be probed by space-borne interferometers LISA \cite{LISAConsortiumWaveformWorkingGroup:2023arg}, Taiji \cite{Luo:2021qji}, TianQin \cite{Luo:2025ewp} and some other GW detectors. In this section, we study the induced GW induced by the enhanced curvature perturbation in hybrid inflation during radiation-dominated epoch. Following \cite{Pi:2020otn,Domenech:2021ztg}, it is
\begin{align}\label{eq:OmegaGW}
    \Omega_\mathrm{GW,0}(f)h^2&=1.62\times10^{-5}\left(\frac{\Omega_{r,0}h^2}{4.18\times10^{-5}}\right)\left(\frac{g_\ast(f)}{106.75}\right)\left(\frac{g_{\ast,s}(f)}{106.75}\right)^{-4/3}\Omega_\mathrm{GW,r}(f),\\
    \Omega_\mathrm{GW,r}(k)&\equiv\Omega(\eta\to\infty,k)=3\int_0^{\infty}\text{d}v\int_{|1-v|}^{1+v}\text{d}u\mathcal{T}(u,v)\mathcal{P}_\mathcal{R}(ku)\mathcal{P}_\mathcal{R}(kv),\\
    \mathcal{T}(u,v)&=\frac{1}{4}\left(\frac{4v^2-(1+v^2-u^2)^2}{4uv}\right)^2\left[\frac{ (u^2+v^2-3)}{4u^3v^3}\right]^2\nonumber\\
    &\times \bigg{\{}\left[-4uv+(u^2+v^2-3)\ln\bigg{|}\frac{3-(u+v)^2}{3-(u-v)^2}\bigg{|}\right]^2\nonumber\\
      & +\left[\pi(u^2+v^2-3)\Theta_{\frac{1}{2}}(u+v-\sqrt{3})\right]^2\bigg{\}},
\end{align}
where $ \Omega_{r,0}h^2\approx 4.18 \times 10^{-5}$ is the density fraction of radiation today \cite{Planck:2018jri}. $g_\ast$ and $g_{\ast,s}$ are the effective number of relativistic degrees of freedom contributing to the energy density and entropy density, respectively \cite{Husdal:2016haj}, evaluated at the horizon reentry. $f \approx 1.5\times10^{-9}k/\text{(1pc)}^{-1}\text{Hz}$ is the frequency corresponding to the wavenumber $k$.

\begin{figure}[t]
\centering
    \includegraphics[width=0.75\textwidth]{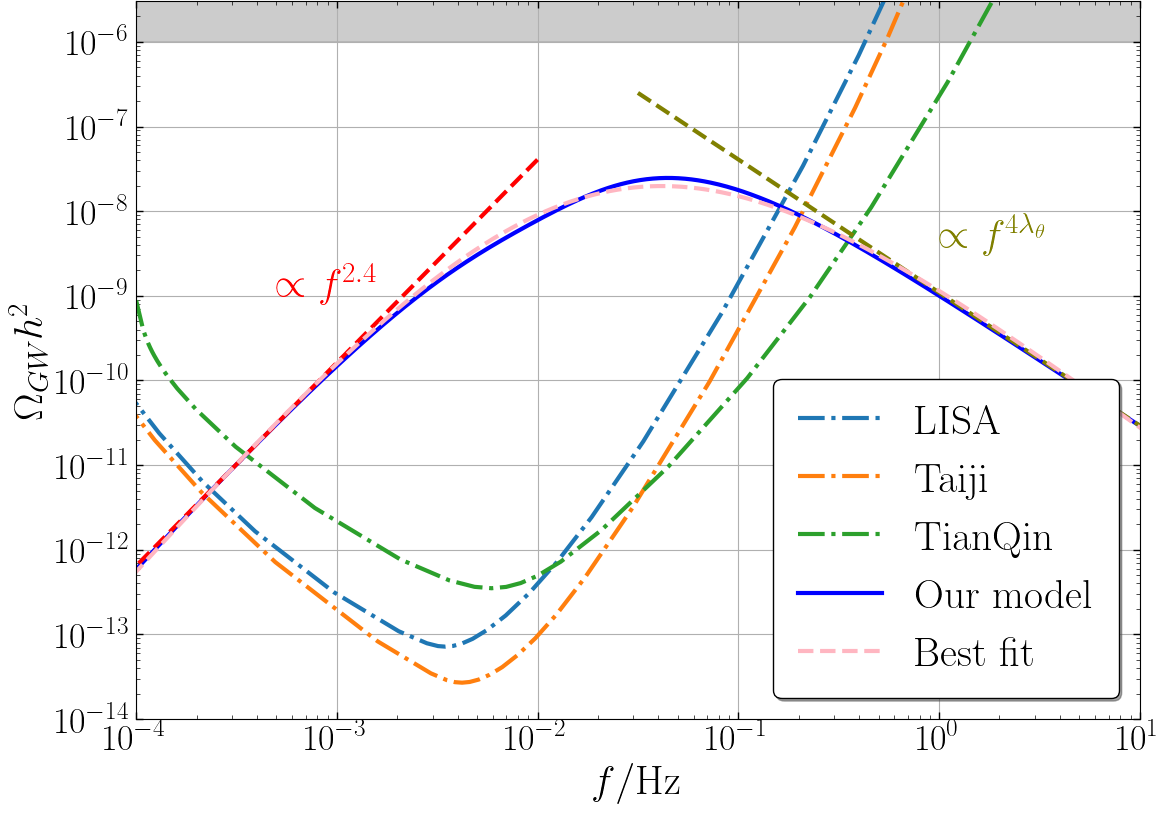}
  \caption{The induced GW spectrum in the current epoch as a function of frequency. The red and yellow dashed lines indicate the power-law behavior of the GW power spectrum in the infrared and ultraviolet regimes; the dot-dashed lines represent the power-law integrated sensitivity curves for LISA, Taiji, and TianQin, assuming an observation time of 3 years and a signal-to-noise ratio of 3;  pink dashed line shows the best-fit result of the GW power spectrum, and the fitting function is given by \eqref{Ffit}. The gray-shaded region is the bound from relativistic DOF from CMB/BBN \cite{Cyburt:2004yc,Binetruy:2012ze,Arbey:2021ysg,Grohs:2023voo}. Parameters are the same as in curvature perturbation power spectrum \eqref{eq:fitPR}.
  }\label{GW}
\end{figure}

The power spectrum $\mathcal{P}_\mathcal{R}(k)$ can be approximated by the semi-analytical formula \eqref{powerP}, 
\begin{align}\label{eq:fitPR}
\mathcal{P}_\mathcal{R}(k)=A_\mathcal{R}\left(\frac{k}{k_c}\right)^3\left[1+B\left(\frac{k}{k_c}\right)^{\frac{1}{\Delta}}\right]^{\Delta(-3+2\lambda_\theta)},
\end{align}
which gives a broken-power-law power spectrum. Here, the parameters are chosen as $A_\mathcal{R} = 2.83, \; B = 2.02, \; \Delta = 0.934, \; \lambda_\theta = 3/2-\sqrt{43/12}\approx-0.39, k_c=1.56\times10^{13}\text{Mpc}^{-1}.$
The GW spectrum in \autoref{GW} can be calculated by either numerical integration, or by using the semi-analytical formula proposed in \cite{Li:2024lxx}. It exhibits a scaling of $\propto f^{2.4}$ in the near-infrared regime and $\propto f^{4\lambda_\theta}$ in the ultraviolet regime. 
In the future, the ultraviolet tilt $n_T$ of the GW spectrum $\Omega_\mathrm{GW}$ can be accurately detected. As we commented, the attractor characteristic root $\lambda_\theta$ is one quarter of $n_T$, which in turn determines the nonlinear parameter $f_\mathrm{NL}\approx-5\lambda_\theta/6$ of the logarithmic non-Gaussianity of the curvature perturbation around the peak \cite{Atal:2019erb,Taoso:2021uvl,Atal:2021jyo}. This is essential in calculating both the PBH abundance and the induced GW spectrum, which can in principle be used to distinguish models.

Similarly, we can parametrize the gravitational-wave power spectrum $\Omega_{\text{GW,0}}(f)h^2$ by a broken-power-law form \cite{Pritchard:2024vix},
which captures the characteristic infrared, intermediate, and ultraviolet scaling behaviors,
\begin{align}
\Omega_{\text{GW,0}}(f) h^2= A_{\rm GW} \frac{(\tilde{\alpha} + \tau)^{\lambda}}
{\left[ \tau \left( \frac{f}{f_0} \right)^{-\tilde{\alpha} /\lambda} 
   + \tilde{\alpha}  \left( \frac{f}{f_0} \right)^{\tau/\lambda} \right]^{\lambda}}.\label{Ffit}
\end{align}
Here the best-fit data are $A_{\rm GW}=1.98\times 10^{-8}$, $f_0=4.16\times 10^{-2}~\rm Hz$, $\tilde{\alpha}=2.68$, $\tau=1.76$, $\lambda=6.28$.
We see that the infrared slope is smaller than the universal scaling $k^3$ \cite{Cai:2019cdl}, as the frequency band we fit is not infrared enough. The fitted curve is slightly larger before the peak, and slightly smaller at the peak, originating from the resonance structure which is more explicit for narrow-peak spectra \cite{Kohri:2018awv,Pi:2020otn}. In the UV tail, \eqref{Ffit} has a good fit of $\Omega_\mathrm{GW}\sim f^{-1.76}$, compared with the semi-analytical result $\Omega_\mathrm{GW}\propto f^{4\lambda_\theta}\sim f^{-1.57}$.

%%%%%%%%%%%%%%%%%%%%%%%%%%%%%%%%%%%%%%%%%%%%%%%%%%%%%% %%%%%%%%%%%%%%%%%%%%%%%%%%%%%%%%%%%%%%%%%%%%%%%%%%%%%% %%%%%%%%%%%%%%%%%%%%%%%%%%%%%%%%%%%%%%%%%%%%%%%%%%%%%%

\section{Conclusion}

In this work, we robustly tracked the superhorizon evolution of multi-field inflationary perturbations by applying the $\delta N$ formalism within an $n$-dimensional parameter space formed by $N$ and its transverse subspace.
We demonstrated that specifying the final hypersurface via a constant single-field value (e.g., $\delta\varphi^1=0$) is valid only when the background trajectory asymptotically aligns with that specific field direction before the end of inflation. For generic multi-field dynamics, incorporating the transverse directions (which encode the isocurvature modes) is essential to accurately capture the late-time evolution of the comoving curvature perturbation.

Applying this framework to two-field hybrid inflation, we identify an amplification mechanism for curvature perturbation that is intrinsically multi-field and qualitatively distinct from single-field non-attractor scenarios. Unlike ultra-slow-roll inflation, which relies on a drastic reduction in the inflaton velocity, the enhancement here is triggered by the tachyonic instability of the waterfall field. This instability causes a rapid exponential growth of the isocurvature perturbations just before the inflationary trajectory turns. Consequently, these growing isocurvature modes efficiently source the adiabatic curvature perturbation around the turn, achieving significant amplification without requiring a sharp deceleration of the background expansion. 

%Applying this framework to two-field hybrid inflation, we uncovered a curvature amplification mechanism that is intrinsically multi-field and qualitatively distinct from single-field non-attractor scenarios. Rather than relying on a transient, drastic reduction in the inflaton velocity (such as in ultra-slow-roll), the enhancement in hybrid inflation is driven by a tachyonic instability of the isocurvature mode associated with the waterfall field. Crucially, this instability develops just prior to the trajectory's turn in field space, allowing the rapidly growing isocurvature mode to efficiently source the adiabatic curvature perturbation without requiring the background expansion to sharply decelerate.

By deriving the exact nonlinear expressions for the curvature perturbation, we revealed that a logarithmic dependence on the waterfall field fluctuation dominates near the spectral peak. This nonlinear relation generates non-Gaussian curvature perturbation that closely mirrors constant-roll
single-field predictions, demonstrating that the waterfall field dynamics predominantly govern both the spectral enhancement and the non-Gaussian distribution of the curvature perturbation at these scales. Importantly, the non-Gaussianity is not an independent parameter in a globally realistic hybrid inflation model: the sign of $f_\mathrm{NL}$ (thus the skewness) is fixed by the tachyonic waterfall geometry, which is always positive, while its amplitude is set by the initial value of the waterfall field and the remaining duration of inflation after the transition. An upper bound of $f_\mathrm{NL}$ is given to avoid quantum diffusion.

Utilizing this analytical framework, we computed the amplified small-scale power spectrum and its corresponding observational signatures. We confirmed that this mechanism can efficiently seed primordial black hole (PBH) formation and generate a stochastic background of induced gravitational waves. As an example, we show the mass function of asteroid-mass PBHs which account for all the dark matter, and the induced GW spectrum at millihertz which can be probed by space-based interferometers in the future.

Our current analysis relies on the assumption of attractor solutions, effectively neglecting independent velocity perturbations within the phase-space $\delta N$ formulation. Extending this formalism to encompass the full set of phase-space degrees of freedom to capture potential transient non-attractor dynamics represents a compelling direction for future work.

\begin{acknowledgments}
We thank Atsushi Naruko and Misao Sasaki for valuable comments. This work is supported by the National Key Research and Development Program of China Grant No. 2021YFC2203004, by the National Natural Science Foundation of China (NSFC) Grants Nos. 12475066 and 12447101, and by JSPS KAKENHI grant No. JP24K00624.
\end{acknowledgments}

\bibliographystyle{apsrev4-2}
\bibliography{multifield}

\end{document}